\documentclass{article}
\usepackage{graphicx}
\pdfoutput=1
\usepackage{amsmath}
\usepackage{amssymb}
\usepackage{mathrsfs}
\usepackage{float}
\usepackage[title]{appendix}
\usepackage[a4paper, total={6in, 9in}]{geometry}
\usepackage[final]{pdfpages}  
\usepackage{comment}
\usepackage{authblk}

\newcommand{\half}{\frac{1}{2}}
\newcommand{\abs}[1]{\left| #1 \right|}
\newcommand{\ket}[1]{\left| #1 \right>}
\newcommand{\bra}[1]{\left< #1 \right|}
\newcommand{\braket}[1]{\left\langle #1 \right\rangle}

\newcommand{\round}[1]{\left( #1 \right)}
\newcommand{\elc}{\xi} % elc stands for this is electron-light-coupling
\newcommand{\elcth}{\frac{\xi\tau}{\hbar}} % elcth stands for this is electron-light-coupling with time over hbar
\newcommand{\E}{\mathscr{E}}
\newcommand{\ktl}{{\abs{k}}}
\newcommand{\kappn}{s}

\title{Supplementary Material: Entanglements of electrons and cavity-photons in the strong coupling regime}
\author{Ofer Kfir}
\date{}
\affil[1]{University of Göttingen, IV. Physical Institute, Göttingen, Germany }

\begin{document}



# Entanglements of electrons and cavity-photons in the strong coupling regime

Ofer Kfir

*University of Göttingen, IV. Physical Institute, Göttingen 37077, Germany*

[ofer.kfir@phys.uni-goettingen.de](ofer.kfir@phys.uni-goettingen.de)

Abstract:

This work sets a road-map towards an experimental realization of strong coupling between free-electrons and photons, and analytically explores entanglement phenomena that emerge in this regime. The proposed model unifies the strong-coupling predictions with known electron-photon interactions. Additionally, this work predicts a non-Columbic entanglement between freely propagating electrons. Since strong-coupling can map entanglements between photon pairs onto photon-electron pairs, it may harness electron beams for quantum communication, thus far exclusive to photons.

Entanglement between the states of propagating quantum entities, and its violation of local realism [1–3], acquires significance technological importance for quantum technologies, such as quantum computation and quantum key distribution [4]. Photon pairs' polarization, expressed as Bell-states, allow for the transport of quantum entanglement over ever-growing distances, under the sea [5] and in the outer space [6]. Matter beams may offer an alternative carrier of quantum information [7,8], which utilizes a richer set of properties, such as shorter wavelength, or the controllability with electric and magnetic fields. Specifically, Fundamental particles, such as free electrons, are especially relevant for entanglement transport. Unlike atoms and other composite matter, free electrons are decoupled from free-space radiation, and they lack an internal structure to decohere into. However, the quantum control of fundamental-particle beams is at its infancy.

Two important phenomena address the energy-exchange of light and electron beams: PINEM and EELS. In PINEM (photo-induced near-field electron microscopy [9]), a strong (and hence classical) laser field accelerates or decelerates electrons in a beam. This concept allows for optical control over the electron quantum wavefunction [10–13], culminating in the prediction [10] and demonstration [14–16] of attosecond-scale electron pulses. Similar effects, such as electron-energy gain spectroscopy (EEGS) [17,18], and effects of the light's ponderomotive-energy, such as the Kapitza-Dirac effect [19] and electron-phase retarders [20] are also well described by classical laser fields. In the abovementioned effects, adding or removing a photon from the laser field makes no difference, since the electron-photon coupling is extremely weak.

In EELS (electron-energy loss spectroscopy) the coupling can be increased by using metallic nanostructures, where the large polarizability and plasmonic resonances [21] allows for detectable signals while the nanometric features compensate for the momentum mismatch between electrons and the optical excitations (i.e. plasmons, see detailed reviews by García de Abajo [22], by Talebi [23], and references within). Such systems were investigated using rings [24], spheres [22], cubes [25,26] and rods [27,28] geometries, as well as for stacked particles [29], ordered or disordered structures [30–32], and also symmetric [33,34] and symmetry-broken systems exhibiting non-Hermitian phenomena [35]. However, rapid decoherence eliminates entanglement features [36] between the electrons and any excited plasmons, stemming from radiative damping [37], intrinsic dissipation [38] absorption and sensitivity to defects. Alternatively, transparent dielectrics have no theoretical bound for the excitation probability [30]. Optical excitations in such dielectrics can be readily injected, collected or manipulated, but unfortunately, their coupling to electron beams is weak.

This article proposes that the phase-matching of swift electrons to photons confined in a fiber-based dielectric cavity can increase the interaction towards the strong coupling regime, and theoretically investigates emergent entanglement phenomena. The phase-matching bandwidth can isolate even a single cavity mode, allowing for a single channel of energy exchange between matter and radiation. An analytical entanglement model is developed for the electron interaction with a cavity mode, which can apply for any coupling regime, weak and strong. To exemplify novel phenomena arising in the strong coupling regime, this work focuses on two scenarios: first, the entanglement between cavity photons and a traversing electron is investigated and compared rigorously to the known phenomena of PINEM and EELS in the limit of weak coupling. Second, the cavity capability to mediate non-Coulombic entanglement between two distant electrons within a beam is explored. Finally, the strength and spectral properties of the coupling are evaluated *quantitatively*, for the case of a stadium-like whispering-gallery mode cavity based on a single-mode fiber.

The proposed experimental concept for reaching strong coupling is shown in Figure 1 (more details are discussed later on and in Figure 4). The optical excitations are whispering gallery modes (WGM) of a stadium-geometry cavity based on a thin waveguide. An electron beam passing parallel to a straight section of the cavity, which is a single mode fiber, excites a bound mode via the interaction with the mode's evanescent tails in the vacuum. The conservation of both energy and momentum between the electron and light is fulfilled only for a specific photon energy, for which the electron travels at the mode's phase velocity. The dispersion in the fiber narrows the interaction's spectral bandwidth. In the example shown on Figure 1b, a silicon-nitride ($Si_3N_4$) cavity is optimized to couple electron beams accelerated to 200 keV with photons having vacuum wavelength near 1.064 μm ($\hbar\omega_0 = 1.1654\ eV$). The coupling bandwidth

narrows down to only 0.011 eV after a propagation length of 100 µm. Importantly, the momentum-matching condition, also referred to as phase-matching, allows for the signal within the interaction bandwidth to grow coherently, and to approach strong coupling.

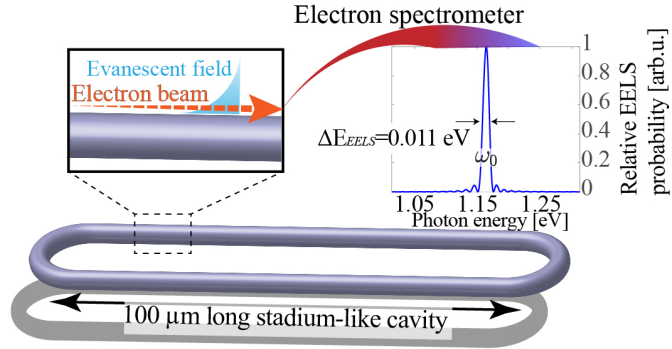

Figure 1 – Proposed experiment for a narrow-bandwidth strong coupling. (a) Evanescent optical field couples a cavity mode to an adjacent electron. (b) Phase matching between the electron and the cavity-photon limits the coupling to a narrow spectral band. For example, a 100 µm propagation near a $Si_3N_4$ cavity limits the coupling bandwidth (expressed as EELS bandwidth) to 11 meV around $\hbar\omega_0 = 1.1654\ eV$ ($\lambda = 1064\ nm$). Further details are in the text.

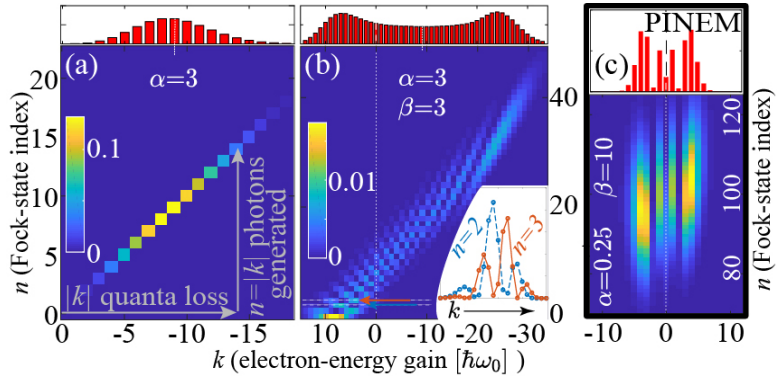

Figure 2 – Electron-photon entanglements patterns. (a) Colormap of $|c_{n,k}|^2$, the co-incident probabilities of photons with electron-energy gain $k\hbar\omega_0$, after a strong interaction ($\alpha = 3$) with an empty cavity. (b) Rich entanglement features for an initial coherent state $|\beta = 3\rangle$, in the cavity. Oscillations in the electron spectra co-incident with Fock-states, c.f. inset for $n = 2$ and $n = 3$, are absent from the integrated electron spectrum (top axis, red bars) since Fock states are orthogonal. (c) The known electron spectrum for PINEM, $|c_k|^2 = |J_k(2|g|)|^2$ (top, red bars), emerges for weak coupling and highly-populated cavity ($\alpha = 0.25$, $|\beta\rangle = |10\rangle$), with $|g| = |\alpha\beta|$. In this limit the electrons and photon-states separable.

Coincident probability, $|c_{n,k}|^2$, for $n$ photons and electron-energy gain of $k\hbar\omega_0$, for an electron ultra-strongly coupled ($\alpha = 3$) to an empty cavity. (b) same as (a), but with a coherent-state in the cavity, $|\beta = 3\rangle$, results in rich entanglement patterns. The inset shows electron spectral oscillations for co-incident detection with Fock-states $n = 2$ and $n = 3$. The overall electron spectrum (top axis, red bars) is smooth since electron spectra co-incident with different $n$ add incoherently. (c) The known electron spectrum for PINEM, $|c_k|^2 = |J_k(2|g|)|^2$ (top, red bars), emerges for weak coupling and highly-populated cavity ($\alpha = 0.25, |\beta\rangle = |10\rangle$), with $|g| = |\alpha\beta|$. In this regime the electrons and cavity photons are nearly independent, and hence separable.

The analytical model relays on the narrow spectral response of the proposed system. Only photons having angular frequency $\omega_0$ are phase-matched with the relativistic electron beam. The narrow-bandwidth suppresses finite electron wavepacket [39,40] effects. The quantum state of these two systems can be described as energy-ladder systems with $\hbar\omega_0$ spacing between the levels: The photon Fock-states in the cavity are represented by $|n\rangle$, a semi-infinite ladder with $n \geq 0$. The electron states, $|E_k\rangle$, with $E_k = E_0 + k\hbar\omega_0$, represent gain ($k > 0$) and loss ($k < 0$) with respect to the "zero-loss energy", $E_0$. $\hbar$ is the reduced Plank's constant. Thus, a general state of such electron-photon system can be written as

$$|\psi\rangle = \sum_{n=0}^{\infty} \sum_{k=-\infty}^{\infty} c_{n,k} |E_k, n\rangle. \tag{1}$$

The relation between the state of the system before and after the interaction can be described by the scattering matrix, $\hat{S}$, as $|\psi_{final}\rangle = \hat{S}|\psi_{initial}\rangle$. Neglecting electron dispersion effects allows to write $\hat{S}$ as exchanging energy between the electrons and the photons,

$$\hat{S} = D(\hat{b}\alpha) = e^{\alpha\hat{b}\hat{a}^\dagger - \alpha^*\hat{b}^\dagger\hat{a}}. \tag{2}$$

Here, $\alpha$ is the coupling strength, $\hat{a}, \hat{a}^\dagger$ are the non-commuting photon ladder operators, and $\hat{b}, \hat{b}^\dagger$ are the commuting electron-energy ladder operators. The commutation, $[\hat{b}, \hat{b}^\dagger] = 0$, result in an algebra similar to scalars, so $\hat{S}$ behaves as a the displacement operator $D(\alpha)$ [41]. A comprehensive treatment of $\hat{b}, \hat{b}^\dagger$, and electron dispersion effects can be found in Sections S.4 and S.1 of the supplementary, respectively.

The interaction of a relativistic electron with an empty cavity is an important and instructive case to consider (see Figure 2a). The state of the combined electron-photon system before any interaction can be written as a pure state, with the electron at the zero-loss energy and no photons.

$$|\psi_i\rangle = |E_0, 0\rangle. \tag{3}$$

As the interaction is a displacement operator, the final state after the first interaction, $|\psi_f\rangle$, is a coherent state [42], as for plasmons [43], in which energy conservation entangles each optical Fock-state to an equal electron-energy loss, $E_k = E_{-n}$.

$$|\psi_f\rangle = \sum_{n \geq 0}^{\infty} e^{-\frac{|\alpha|^2}{2}} \frac{\alpha^n}{\sqrt{n!}} |E_{-n}, n\rangle. \tag{4}$$

One can consider eq. (4) as the multi-level electron-photon equivalent of a Bell-pair, $|\psi_f\rangle = (c_0|E_0, 0\rangle + c_1|E_{-1}, 1\rangle + \cdots)$. Thus, coincidence measurements are expected to expose correlations between the measured electron-energy loss and photon detection. For comparison with EELS experiments, one should consider a weak coupling, where only one energy-loss channel is detectable, with population probability of $|\alpha|^2$. Higher EELS orders [43] necessitates a strong coupling $|\alpha| \sim 1$. In the supplementary, S.2, strong coupling EELS is derived as "field-less PINEM" to touch upon their equivalence. A general feature of the electron-energy distribution is that the average loss is $E_0 - \langle E \rangle = |\alpha|^2$, in either weak or strong coupling. The one-to-one entanglement between the two entities is expressed as the purely diagonal electron-photon energy-state population in Figure 2a.

For a quantum-optics description of PINEM experiments one needs to consider a coherent state $|\beta\rangle$ with an average number of $|\beta|^2$ photons in the cavity. In this case, the exact final electron-photon quantum state is characterized by eq. (1) with the coefficients $c_{n,k}^{PINEM} = \langle E_k, n|D(\hat{b}\alpha)|E_0, \beta\rangle$,

$$c_{n,k}^{PINEM} = e^{\frac{|\alpha|^2 - |\beta|^2}{2}} \frac{\beta^{(n+k)} \alpha^k}{\sqrt{n!}} \sum_{\ell=0}^{\infty} \left[ \frac{(n+k+\ell)!}{(n+k)!} \right] \frac{(-|\alpha|^2)^\ell}{(k+\ell)!\,\ell!}. \tag{5}$$

The supplementary section S.2 details the algebraic derivation. Figure 2b presents the electron-photon spectral probability map for the case of a strong coupling ($\alpha = 3$) to a cavity populated with 9 photons in average, $\beta = 3$. The entanglement correlates diagonally as for an empty cavity, but also includes rich patterns. Specifically, the electron spectrum depends strongly on the coincident Fock state (see inset for $n = 2$ and $n = 3$). The photon-averaged electron spectra is smooth (red bars, top axis), similar to a spatial scattering of atoms off a coherent photon state [44,45]. Consistently with EELS and with the case of an empty cavity, the mean energy-loss is $|\alpha|^2$. The electron energy distribute nearly symmetrically around the mean, with a spectral widths of $4|\alpha\beta|$. This generalizes PINEM [9,46,47], known for its $4|g|$ bandwidth, symmetric around $E_0$ (special case of $|\alpha|^2 \to 0$) [10]. $g$ is the PINEM Rabi-parameter.

To exactly retrieve known PINEM spectra, with probability amplitudes $c_k = J_k(2|g|)$ [10,46], one needs to consider strong optical fields $|\beta| \gg 1$, weakly coupled to the electron beam, $\alpha \ll 1$ (see detailed calculation in the supplementary section S.2.2.1). $J_k$ is the Bessel function of the first kind. Eq. (5) reduces

to Bessel-function amplitudes when approximating the square brackets within it as $\frac{(n+k+\ell)!}{(n+k)!} \approx (n+k)^\ell$.

Thus, one can write $(n+k)^\ell(-|\alpha|^2)^\ell = \left(-\left|\alpha\sqrt{(n+k)}\right|^2\right)^\ell$. The summation then transforms to $\left(\sqrt{(n+k)}\right)^{-k} e^{ik \arg g} \cdot J_k(2|g|)$, with the definition.

$$g = \alpha\sqrt{(n+k)} \approx \alpha|\beta|. \tag{6}$$

Neglecting quantum fluctuations added to the light by the interaction, $\left(\sqrt{(n+k)}\right)^{-k} = \left(\sqrt{\langle n+k \rangle}\right)^{-k}$, decouples the electron and photon states. Thus, the decoupled state

$$|\psi_{PINEM}\rangle = |\beta\rangle \otimes J_k(2|g|)|E_k\rangle, \tag{7}$$

fully retrieves the known PINEM Bessel-amplitudes decoupled from the driving laser field. This decoupling allows to consider the electron *wavefunction*, and its laser-modulation. Furthermore, the Rabi-parameter, $g$ emerges naturally as the product of the coherent state amplitude and the coupling strength. Figure 2c shows the electron-photon spectra for $\beta = 10, \alpha = 0.25$, that is, a cavity with 100 photons in average, weakly coupled to the electron beam. The electron spectrum is nearly independent of the photon state, yielding the electron spectral oscillations typical for Bessel amplitudes (see zoomed electron spectrum).

The second strong-coupling phenomena exemplified here is non-Coulombic entanglement of two consecutive electrons in a beam, mediated by long lived cavity photons. A lifetime of 10 picoseconds allows excitations of the first electron to affect the second, while suppressing Coulombic interactions between them. The passage of the first electron generates a state as in eq. (4). A second electron with an equal zero-loss energy (marked here $\mathcal{E}_0$ to distinct from the first electron) will result in a three-particle state, $|\psi_f^{e-e}\rangle \sum_{n=0}^\infty \sum_{k=-\infty}^n c_{n,k}^{e-e}|E_{-n}, \mathcal{E}_k, n-k\rangle$, characterized by 2-indices $c_{n,k}^{e-e} = \langle E_{-n}, \mathcal{E}_k, m|D(\hat{b}\alpha)|E_{-n}, \mathcal{E}_0, n\rangle$. $k$ is the energy quanta gained by the second electron and $n$ is the Fock-state index prior to the arrival of the second electron, which is also the final energy state of the first electron $|E_{-n}\rangle$. The final Fock-state of the cavity is $|n-k\rangle$. Thus,

$$c_{n,k}^{e-e} = \frac{\alpha^n \alpha^k}{n!\sqrt{(n-k)!}} \sum_{\ell=0}^\infty (n+\ell)! \frac{(-|\alpha|^2)^\ell}{(k+\ell)!\,\ell!}. \tag{8}$$

Explicit derivation are in the supplementary section S.3. One can think of such an event as PINEM, pumped by the first electron, as apparent in the similarity of eqs. (5) and (8). Figure 3 shows the resulting entanglement features for strong couplings of $\alpha = 1$ and $\alpha = 3$. The single particle spectra (right axis, red bars) is smooth, while oscillations appear in co-incidence measurements. See the inset of Figure 3b for a

spectrum of the 2nd electron, co-incident with a 12-quanta-loss state of the first electron. In such an electron-pair experiment, energy-gains are unique to the second electron, and hence could be used to record co-incident spectra without separating the paths of the two electrons.

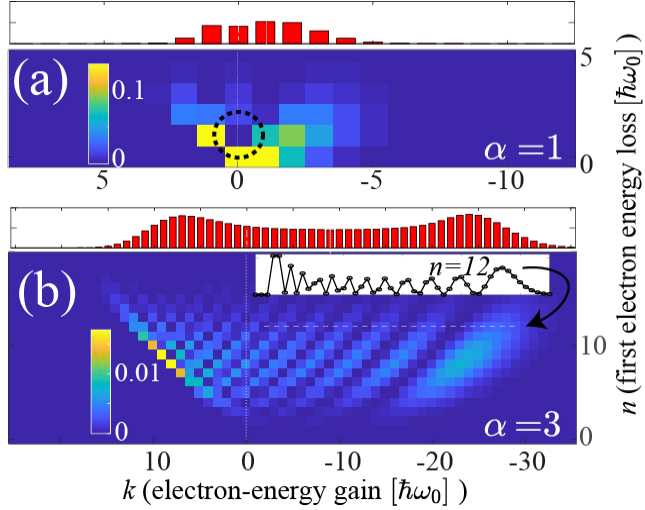

Figure 3 – Electron-electron interaction for two distant electrons in a beam, mediated by long-lived photons. The colormap, $|c_{n,k}^{e-e}|^2$, is the co-incident probability for $n\hbar\omega_0$ energy-loss of the first electron and $k\hbar\omega_0$ gain of the second. (a) Strong coupling, $\alpha = 1$, allows for mutually exclusive states (dash circle) of the electron pair, where if the first electron loses one quantum, the second cannot be loss-less. (b) Stronger interactions induces rich entanglement features (see inset for lineout of the second electron spectra, co-incident with $n = 12$).

The last part this letter utilizes the above derivations to *quantitatively* evaluate the coupling constant via PINEM. Specifically, the term $g = \alpha|\beta|$ in eq. (6) links the coupling constant to the classical acceleration of an electron by the mode's field. The field is represented as a coherent state $|\beta\rangle$, and the acceleration or deceleration is represented by $g$. For an electron traveling the path $0 < z < L$ near a straight arm of the cavity, the parallel acceleration is given by the light-field component, $E_z^{(\omega)}(z, t(z))$, evaluated for time $t(z)$. The superscript $\omega$ indicates a frequency dependence. The classically calculated electron-energy gain within the interaction region is $q \int_0^L E_z^{(\omega)}(z, t(z))dz$, where $q$ is the electron charge (transverse recoil is negligible, see supplementary section S.1.2.2). $g$ is then the unit-less ratio between the electron-energy gain and the photon energy (eq. (3) in ref. [10]), which with eq. (6) determines the coupling constant as

$$\alpha = \frac{g}{|\beta|} = \frac{1}{|\beta|}\frac{q}{2\hbar\omega}\int_0^L E_z^{(\omega)}(z, t(z))dz. \tag{9}$$

Although $\alpha$ is calculated from classical fields, it is a geometrical property of the apparatus for a given electron zero-loss energy, since $|\beta| = \sqrt{\langle n \rangle} \propto E_z$. Intuitively, one can interpret $\alpha$ as the strength of a PINEM effect for one cavity photon. Eq. (9) has few important aspects, especially when implemented to a

long interaction length, e.g. many μm: (i) The optimal coupling occurs when $E_z^{(\omega)}(z, t(z))$ is constant along the electron trajectory, which occurs at perfect phase-matching.(ii) Conceptually, the ultimate coupling would be for a straight fiber with periodic boundary conditions and length $L$. One can realistically reach $1/\sqrt{2}$ of that, when accounting for the backward propagating mode in a realistic cavity. (iii) The coupling bandwidth is limited by dispersion (see Figure 1b).

A cavity design for a strong coupling requires a small optical-mode volume, matching the velocity of the electron with that of the mode, and a meaningful field component $E_z$ in vacuum. Those can be achieved in WGM cavity, based on a straight single-mode fiber (see Figure 1a). The small fiber diameter plays multiple roles: it minimizes the mode volume, increases the evanescent tails in vacuum that interact with the electron, increase the field component, $E_z$, and pushes the modal phase velocity towards the speed of the relativistic electrons. The stadium-geometry of the proposed resonator allows for a long straight section, while minimizing additional mode volume and losses.

Figure 4 shows the coupling of electron beams to a mode with a photon energy of $\hbar\omega_0 = 1.1654\ eV$ (vacuum-wavelength $\lambda = 1064\ nm$) in a step index profile [48]. For the selected parameters (fiber diameter of 463 nm, electron zero-loss energy of 200 keV) the coupling reaches $\alpha = 0.76$ for 100 μm interaction length. Figure 4b presents the spectral width of the interaction, derived from the coherence length [49]. Comparison to the free spectral range for cavities with total circumference of $2L$ (dashed line), indicates the number of modes with which the electron may couple. A $Si_3N_4$ cavity (orange line) allows for phase-matching with 4 optical modes, while silicon (blue line, for 213 nm diameter), allows for the coupling of a single optical mode, or none.

As a concrete example, I consider the stadium-shaped WGM cavity with a 100 μm straight arm and a negligible semicircle circumference, comprising a $Si_3N_4$ fiber with a diameter of 463 nm. An electron beam passing close to the cavity wall would have a coupling strength of $\alpha = 0.76/\sqrt{2} = 0.53$, where the $\sqrt{2}$ accounts for the non-interacting cavity-arm. The field evanescence length 120 nm in vacuum. Thus, an electron beam with a semiconvergence angle of 0.15 mili-radians and a waist of 10 nm would experience a uniform light-field. These are achievable parameters in contemporary electron microscopes. With this design scheme, and possible other modifications, unprecedented strong coupling and long-lived entanglement effects may be reached in the not-distant future.

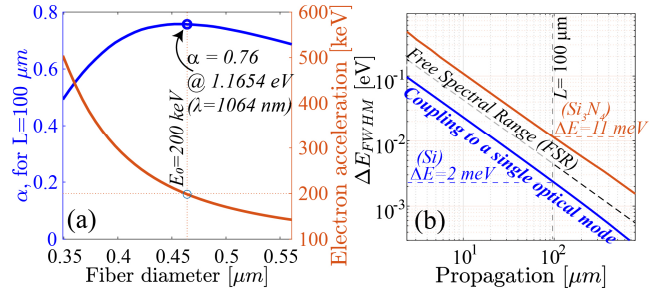

Figure 4 – (a) The coupling constant to a 100-µm-long $Si_3N_4$ step-index fiber (left axis, blue) and the phase-matched electron energy (right axis, orange) for 1064 nm photons, as a function of the fiber diameter. This calculation assumes periodic boundary conditions, thus, a realistic coupling would be smaller by $\sqrt{2}$. (b) The coupling bandwidth (orange) narrows for longer interactions. Increased dispersion (e.g. Si, blue line) can result in a narrower bandwidth than the free spectral range (dash diagonal), which limits the coupling to a single optical mode at most.

To conclude, this work proposes a path towards a strong coupling regime between electrons and cavity-photons based on narrow-band phase-matching, and investigates phenomena that this regime may enable. The analytical model addresses EELS and PINEM on an equal footing alongside strong coupling phenomena. The coupling, $\alpha$, may be retrieved experimentally using PINEM, via eq. (9), or using EELS, via the $|\alpha|^2$ loss probability, per optical mode. Additionally, strong coupling to a cavity can entwine the quantum state of two consecutive electrons, entangling their final energies. These phenomena, and the concrete design approach brought here set a road-map for experiments of free-electrons strongly-coupled with photons. In the future, the ability to imprint quantum-optical states on relativistic electron-beams may enable the use of electrons as information carriers. The fundamental differences between light and electrons may open new horizons. One example is the use of the high efficiency of electron-detection to herald single- or multiple-photon sources. Another is long-distance communication in space, where matter beams exhibit superior divergence properties, and allow for manipulation with electric and magnetic fields.

## References

1. A. Einstein, B. Podolsky, and N. Rosen, "Can Quantum-Mechanical Description of Physical Reality Be Considered Complete?," Phys. Rev. **47**, 777–780 (1935).
2. G. Weihs, T. Jennewein, C. Simon, H. Weinfurter, and A. Zeilinger, "Violation of Bell's Inequality under Strict Einstein Locality Conditions," Phys. Rev. Lett. **81**, 5039–5043 (1998).
3. A. Aspect, "Bell's inequality test: more ideal than ever," Nature **398**, 189 (1999).
4. M. A. Nielsen and I. L. Chuang, *Quantum Computation and Quantum Information*, 10th anniversary ed (Cambridge University Press, 2010).
5. S. Wengerowsky, S. K. Joshi, F. Steinlechner, J. R. Zichi, S. M. Dobrovolskiy, R. van der Molen, J. W. N. Los, V. Zwiller, M. A. M. Versteegh, A. Mura, D. Calonico, M. Inguscio, H. Hübel, L. Bo, T. Scheidl, A. Zeilinger, A. Xuereb, and R. Ursin, "Entanglement distribution over a 96-km-long submarine optical fiber," Proceedings of the National Academy of Sciences **116**, 6684–6688 (2019).

6. J. Yin, Y. Cao, Y.-H. Li, J.-G. Ren, S.-K. Liao, L. Zhang, W.-Q. Cai, W.-Y. Liu, B. Li, H. Dai, M. Li, Y.-M. Huang, L. Deng, L. Li, Q. Zhang, N.-L. Liu, Y.-A. Chen, C.-Y. Lu, R. Shu, C.-Z. Peng, J.-Y. Wang, and J.-W. Pan, "Satellite-to-Ground Entanglement-Based Quantum Key Distribution," Physical Review Letters **119**, 200501 (2017).
7. L.-M. Duan, A. Sørensen, J. I. Cirac, and P. Zoller, "Squeezing and Entanglement of Atomic Beams," Phys. Rev. Lett. **85**, 3991–3994 (2000).
8. P. Berg, S. Abend, G. Tackmann, C. Schubert, E. Giese, W. P. Schleich, F. A. Narducci, W. Ertmer, and E. M. Rasel, "Composite-Light-Pulse Technique for High-Precision Atom Interferometry," Phys. Rev. Lett. **114**, 063002 (2015).
9. B. Barwick, D. J. Flannigan, and A. H. Zewail, "Photon-induced near-field electron microscopy," Nature **462**, 902–906 (2009).
10. A. Feist, K. E. Echternkamp, J. Schauss, S. V. Yalunin, S. Schäfer, and C. Ropers, "Quantum coherent optical phase modulation in an ultrafast transmission electron microscope," Nature **521**, 200–203 (2015).
11. C. Kealhofer, W. Schneider, D. Ehberger, A. Ryabov, F. Krausz, and P. Baum, "All-optical control and metrology of electron pulses," Science **352**, 429–433 (2016).
12. K. E. Echternkamp, A. Feist, S. Schäfer, and C. Ropers, "Ramsey-type phase control of free-electron beams," Nature Physics **12**, 1000–1004 (2016).
13. G. M. Vanacore, I. Madan, G. Berruto, K. Wang, E. Pomarico, R. J. Lamb, D. McGrouther, I. Kaminer, B. Barwick, F. J. G. de Abajo, and F. Carbone, "Attosecond coherent control of free-electron wave functions using semi-infinite light fields," Nature Communications **9**, 2694 (2018).
14. K. E. Priebe, C. Rathje, S. V. Yalunin, T. Hohage, A. Feist, S. Schäfer, and C. Ropers, "Attosecond electron pulse trains and quantum state reconstruction in ultrafast transmission electron microscopy," Nature Photonics **11**, 793–797 (2017).
15. Y. Morimoto and P. Baum, "Diffraction and microscopy with attosecond electron pulse trains," Nature Physics **14**, 252 (2018).
16. M. Kozák, N. Schönenberger, and P. Hommelhoff, "Ponderomotive Generation and Detection of Attosecond Free-Electron Pulse Trains," Phys. Rev. Lett. **120**, 103203 (2018).
17. F. J. García de Abajo and M. Kociak, "Electron energy-gain spectroscopy," New Journal of Physics **10**, 073035 (2008).
18. P. Das, J. D. Blazit, M. Tencé, L. F. Zagonel, Y. Auad, Y. H. Lee, X. Y. Ling, A. Losquin, C. Colliex, O. Stéphan, F. J. García de Abajo, and M. Kociak, "Stimulated electron energy loss and gain in an electron microscope without a pulsed electron gun," Ultramicroscopy **203**, 44–51 (2019).
19. D. L. Freimund, K. Aflatooni, and H. Batelaan, "Observation of the Kapitza–Dirac effect," Nature **413**, 142–143 (2001).
20. O. Schwartz, J. J. Axelrod, S. L. Campbell, C. Turnbaugh, R. M. Glaeser, and H. Müller, "Laser control of the electron wave function in transmission electron microscopy," arXiv:1812.04596 [physics, physics:quant-ph] (2018).
21. A. Campos, N. Troc, E. Cottancin, M. Pellarin, H.-C. Weissker, J. Lermé, M. Kociak, and M. Hillenkamp, "Plasmonic quantum size effects in silver nanoparticles are dominated by interfaces and local environments," Nature Physics **15**, 275–280 (2019).
22. F. J. García de Abajo, "Optical excitations in electron microscopy," Rev. Mod. Phys. **82**, 209–275 (2010).
23. N. Talebi, "Interaction of electron beams with optical nanostructures and metamaterials: from coherent photon sources towards shaping the wave function," J. Opt. **19**, 103001 (2017).
24. E. J. R. Vesseur, F. J. G. de Abajo, and A. Polman, "Broadband Purcell enhancement in plasmonic ring cavities," Physical Review B **82**, (2010).
25. G. Unger, A. Trügler, and U. Hohenester, "Novel Modal Approximation Scheme for Plasmonic Transmission Problems," Physical Review Letters **121**, (2018).
26. M. J. Lagos, A. Trügler, U. Hohenester, and P. E. Batson, "Mapping vibrational surface and bulk modes in a single nanocube," Nature **543**, 529–532 (2017).

27. S. V. Yalunin, B. Schröder, and C. Ropers, "Theory of electron energy loss near plasmonic wires, nanorods, and cones," Phys. Rev. B **93**, 115408 (2016).
28. A. Hörl, A. Trügler, and U. Hohenester, "Tomography of Particle Plasmon Fields from Electron Energy Loss Spectroscopy," Physical Review Letters **111**, 076801 (2013).
29. G. Haberfehlner, F.-P. Schmidt, G. Schaffernak, A. Hörl, A. Trügler, A. Hohenau, F. Hofer, J. R. Krenn, U. Hohenester, and G. Kothleitner, "3D Imaging of Gap Plasmons in Vertically Coupled Nanoparticles by EELS Tomography," Nano Letters **17**, 6773–6777 (2017).
30. Y. Yang, A. Massuda, C. Roques-Carmes, S. E. Kooi, T. Christensen, S. G. Johnson, J. D. Joannopoulos, O. D. Miller, I. Kaminer, and M. Soljačić, "Maximal spontaneous photon emission and energy loss from free electrons," Nature Physics **14**, 894 (2018).
31. I. Kaminer, S. E. Kooi, R. Shiloh, B. Zhen, Y. Shen, J. J. López, R. Remez, S. A. Skirlo, Y. Yang, J. D. Joannopoulos, A. Arie, and M. Soljačić, "Spectrally and Spatially Resolved Smith-Purcell Radiation in Plasmonic Crystals with Short-Range Disorder," Phys. Rev. X **7**, 011003 (2017).
32. N. Talebi, "A directional, ultrafast and integrated few-photon source utilizing the interaction of electron beams and plasmonic nanoantennas," New J. Phys. **16**, 053021 (2014).
33. S. Guo, N. Talebi, A. Campos, M. Kociak, and P. A. van Aken, "Radiation of Dynamic Toroidal Moments," ACS Photonics **6**, 467–474 (2019).
34. P. Das, H. Lourenço-Martins, L. H. G. Tizei, R. Weil, and M. Kociak, "Nanocross: A Highly Tunable Plasmonic System," J. Phys. Chem. C **121**, 16521–16527 (2017).
35. H. Lourenço-Martins, P. Das, L. H. G. Tizei, R. Weil, and M. Kociak, "Self-hybridization within non-Hermitian localized plasmonic systems," Nature Physics **14**, 360 (2018).
36. P. Schattschneider and S. Löffler, "Entanglement and decoherence in electron microscopy," Ultramicroscopy **190**, 39–44 (2018).
37. D.-L. Hornauer, "Light scattering experiments on silver films of different roughness using surface plasmon excitation," Optics Communications **16**, 76–79 (1976).
38. T. Inagaki, K. Kagami, and E. T. Arakawa, "Photoacoustic observation of nonradiative decay of surface plasmons in silver," Phys. Rev. B **24**, 3644–3646 (1981).
39. A. Gover and Y. Pan, "Dimension-dependent stimulated radiative interaction of a single electron quantum wavepacket," Physics Letters A **382**, 1550–1555 (2018).
40. Y. Pan and A. Gover, "Spontaneous and Stimulated Emissions of Quantum Free-Electron Wavepackets - QED Analysis," arXiv:1805.08210 [physics, physics:quant-ph] (2018).
41. M. O. Scully and M. S. Zubairy, *Quantum Optics* (Cambridge University Press, 1997).
42. R. J. Glauber, "Coherent and Incoherent States of the Radiation Field," Phys. Rev. **131**, 2766–2788 (1963).
43. F. J. García de Abajo, "Multiple Excitation of Confined Graphene Plasmons by Single Free Electrons," ACS Nano **7**, 11409–11419 (2013).
44. V. M. Akulin, F. L. Kien, and W. P. Schleich, "Deflection of atoms by a quantum field," Physical Review A **44**, R1462–R1465 (1991).
45. A. M. Herkommer, V. M. Akulin, and W. P. Schleich, "Quantum demolition measurement of photon statistics by atomic beam deflection," Physical Review Letters **69**, 3298–3301 (1992).
46. S. T. Park, M. Lin, and A. H. Zewail, "Photon-induced near-field electron microscopy (PINEM): theoretical and experimental," New J. Phys. **12**, 123028 (2010).
47. S. T. Park and A. H. Zewail, "Relativistic Effects in Photon-Induced Near Field Electron Microscopy," J. Phys. Chem. A **116**, 11128–11133 (2012).
48. C.-L. Chen, *Foundations for Guided-Wave Optics* (John Wiley & Sons, 2006).
49. R. W. Boyd, *Nonlinear Optics*, 2nd ed. (Academic Press, 2003).



\maketitle

\tableofcontents
\newpage

\section{Basics of the coherent interaction between electronic and photonic states} 

\subsection{Assumptions}
\begin{itemize}
    \item The electron state is .
    \begin{align}
        \left|\psi \right>= \sum_{n=0}^\infty \sum_{j=-\infty}^\infty c_{nj} \left|E_j,n \right>.
    \end{align}
    The electronic part of the state may just be written as $\left|E_j \right>$, similar to Feist et al. 2015 \cite{feist_quantum_2015}.
    \item  The ladder operators are $\hat{b}, \hat{b}^\dagger$, with the commutation relation $\left[\hat{b} ,\,\hat{b}^\dagger\right]= 0$. This commutation relation holds throughout relevant the energy spectrum. For fast electrons,$\round{\hat{b}}^\dagger=\hat{b}^\dagger$. For very slow electrons, the lowering and lifting operators are not the hermitian conjugate of each other. See details in section \ref{section_for_rel_electrons}.
    \item The electrons interact with a harmonic system, with energy spacing $\omega_0$, such that $E_{j+k}=E_j+k\hbar\omega_0$.
    \item For the photons the ladder operators are the standard $\hat{a}$ and $\hat{a}^\dagger$, with $\left[\hat{a},\,\hat{a}^\dagger \right]=1$.(See for example quantization in Scully and Zubairy \cite{scully_quantum_1997} or Mandel and Wolf \cite{Mandel_Wolf_optical_1995})
    \item The operator of the electric field for \textbf{free space}, is 
    \[\vec{\hat{\mathbf{E}}}\left(\mathbf{r},t\right) = \frac{1}{L^{3/2}}\sum_{\mathbf{k}}{\sum_s{\sqrt{\frac{\hbar \omega}{2 \varepsilon_0}}\left[ i\hat{a}_{\mathbf{k},s}\left( 0 \right) \varepsilon_{\mathbf{k},s} e^{i \left(\mathbf{k\cdot r}-\omega t \right) } + h.c. \right]}}
    .\]
    \\
    That can be written for simplicity as \[\vec{\hat{\mathbf{E}}}\left(\mathbf{r},t \right) = \vec{\hat{\mathbf{E}}}^{(+)}\left(\mathbf{r},t\right) +\vec{\hat{\mathbf{E}}}^{(-)}\left(\mathbf{r},t\right)\] 
    Here I just abbreviate 
    \begin{align}
        \vec{\hat{\mathbf{E}}}^{(+)} = \hat{a}
        ,\end{align}
        where the time and space dependency are understood. The quantization of fiber modes has a similar form, see section \ref{section_quantization_of_fiber_modes}.

\item The interaction Hamiltonian is $\mathcal{H}_e+\mathcal{H}_{np}+\mathcal{H}_I$. 

    The Hamiltonian for a local interaction is 
    \begin{align}
        \mathcal{H}=\sqrt{E_{rest}^2+\round{\hat{P}c}^2}+\hbar\omega_0\hat{a}^\dagger \hat{a} + \elc \left(\hat{b}\hat{a}^\dagger+ \hat{b}^\dagger\hat{a}\right).
    \end{align}
    The exact form of the electron Hamiltonian, $\mathcal{H}_e$, and the operators $b,b^\dagger$ is described in section \ref{section_for_rel_electrons}. $\elc$ is the local coupling strength, with units of energy. It is meaningful only in the context of the total interaction strength, as e.g. in eq. \eqref{S_matrix_as_exponent_1}.
    
\end{itemize}

    \subsection{Electron-dispersion effects}
    \subsubsection{Relevant distance for electron-energy dispersion effects}
    The distances that are relevant for dispersion effects can be evaluated from a Taylor expantion of the phase for an electron plane-wave 
\begin{align}
    \phi_{\round{E,z}}&=\phi_{\round{E,0}}+P_{\round{E}}\frac{z}{\hbar} \\
    &=\phi_{\round{E,0}}+\frac{z}{\hbar c}\sqrt{E^2-E_{rest}^2}\\
    &\overset{\round{E=E_0+\Delta E}}{\approx} \phi_{\round{E,0}}+\frac{z}{\hbar c}\round{\sqrt{E_0^2-E_{rest}^2}+\frac{E_0\round{\Delta E}}{\sqrt{E_0^2-E_{rest}^2}}-\frac{E_{rest}^2\round{\Delta E}^2}{2\round{E_0^2-E_{rest}^2}^{3/2}}  }
.\end{align} 
The dispersion can be neglected for short propagation distances, $z \ll z_{dispersion}$, where   
\begin{equation}
z_{dispersion} = \hbar c \cdot \abs{\frac{E_{rest}^2\round{\Delta E}^2}{2\round{E_0^2-E_{rest}^2}^{3/2}}}^{-1} \label{Dispersion_distance_for_relativistic_electrons}
.\end{equation}
For example, electrons at 200 keV, and 11.65 eV bandwidth (10 orders of photons with 1064 nm vacuum-wavelength), $z_{dispersion}\approx 5.3 \,mm$.

\subsubsection{Estimation of the electron recoil (transverse deflection)} 
Since the optical mode has both parallel field ($E_z$ )and transverse field ($E_x, E_y$) components, it can, in principle, deflect the electron and compromise the validity one-dimentional description, as in this work. For the optical-mode parameters described in this work, say in Figure 4 of the manuscript, this deflection is small.
Given a similar field components, as in our case, one can assume for simplicity $|E_x|\approx |E_z|$. During the interaction time $L/v_0$, the transverse deflection can be estimated by the classical impulse is the added transverse momentum $\Delta P_x=q |E_x| (L/v_0)$. $v_0$ is the group velocity of the relativistic electron. 

Thus, the final deflection $\theta_f$ from the induced transverse momentum is
\begin{align}
\theta_f=\frac{\Delta P_x}{P_0}&=\frac{q |E_x| L}{ v_0 P_0 }\\
&\approx \frac{q |E_z| L}{ v_0 P_0 }
.\end{align}
using the phase-matched coupling $\alpha=\frac{q E_z L}{2\hbar\omega_0}$, $\hbar\omega_0=1.1 eV$, $P_0=\gamma m_0 v_0$, and the parameters for 200 keV electrons, $\gamma=1.39$, and $v_0=0.7 c$, one can write
\begin{align}
\theta_f  &\approx  \frac{q |E_z| L}{ v_0 P_0 }
\approx  \alpha\frac{\overbrace{2 \hbar\omega_0}^{2.2 eV}}{ \underbrace{v_0 P_0}_{\gamma m_0 (0.7 c)^2} } 
\approx \alpha \frac{2.2 eV}{338 keV}
\approx \alpha\cdot 6.5e-6 
.\end{align}
The deflection roughly scales with the coupling, and is only few micro-radians. 
Assuming the acceleration is constant and the deflection trajectory is parabolic, this FINAL deflection correspond to a trajectory $x(z)=\frac{\theta_f}{L} \frac{z^2}{2}$, which result in a final deflection $x(z=L) \approx 0.2 \, nm$ for $\alpha=1$ and $L=100\, \mu m$.

\subsection{The displacement operator - S-matrix approach}
The scattering operator (in the interaction picture), $\hat{S}$ need to be accounted for in full, to be valid for strong couplings,
\begin{align}
    \hat{S}=&\mathcal{T}\exp{\left[-\frac{i}{\hbar}\int_{-\infty}^\infty{\elc \left(\hat{b}\hat{a}^\dagger+ \hat{b}^\dagger\hat{a}\right)}\right]}\\
    =&\exp{\left[-i\elcth \left(\hat{b}\hat{a}^\dagger+ \hat{b}^\dagger\hat{a}\right)\right]} \label{S_matrix_as_exponent_1}
.\end{align}
Here, I removed the time-ordering operator, $\mathcal{T}$, since the there is no time dependence for the operator product $\hat{b} \hat{a}^\dagger$ for interaction lengths short enough to suppress dispersion (see eq. \eqref{Dispersion_distance_for_relativistic_electrons}) \[\hat{b}(t) \hat{a}^\dagger(t)=\hat{b} e^{i\omega t} \hat{a}^\dagger e^{-i\omega t}=\hat{b}\hat{a}^\dagger.\]   $\tau$ is an effective interaction duration. Additional phases, such as temporal-delays accumulated by the electron energy-states, can be neglected below the characteristic dispersion distance, as in eq. \eqref{Dispersion_distance_for_relativistic_electrons}.

\subsubsection{Derivation of the S-matrix as a displacement operator}
Eq. \eqref{S_matrix_as_exponent_1} has the form of the displacement operator, 
     \[D\round{\hat{b}\alpha}=\exp\round{\alpha\hat{b}\hat{a}^\dagger-\alpha^*\hat{b}^\dagger\hat{a}}, \]
      with a substitution
      \begin{align}
      -i\elcth&=\alpha\\
      -i\elcth&=-\alpha^*
      .\end{align}
      
     For this to be correct,$\elcth \in \Re eal$. \\
	Aditional comments: 
     \begin{itemize}
         \item The only difference of $D\round{\hat{b}\alpha}$ from $D\round{\alpha}$ is that $\hat{b}$ is an operator. It implies conservation of energy, where every Fock-state is entangled with the corresponding state of an electron energy loss $\ket{E_{j-n},n}$. The splitting of the energy between the electron channel and photon channel resembles a beam-splitter operator, but with $\hat{b}\hat{a}^\dagger$ instead of the $\hat{a}_1\hat{a}_2^\dagger$
         \item If $\elcth \in \Re eal$, than the translation parameter, $\alpha$, is purely imaginary. This means that when the electrons interact with radiation they change momentum, without instantaneous shifts. Time (or propagation) translates momentum difference to modifications of the probability distribution.
     \end{itemize}

\section{Effects for photons interacting with an electron-beam}

\subsection{EELS as field-less PINEM - strong interaction without a driving field}
A strong interaction depends on the coupling parameter, $\alpha$, with the light field acting only as the initial state. Due to  conservation of energy, one expects to end with energy-loss, $E_{-k}$, with $k>0$. The probability amplitude to find an electron in energy $E_{-k}$ can be written as
\begin{align}
     \bra{E_{-k},n} D_{\round{\hat{b}\alpha}}\ket{E_0,0}
    &\overset{BCH}{=}e^{\frac{\abs{\alpha}^2}{2}}\bra{E_{-k},n} e^{-\alpha^*\hat{b}^\dagger \hat{a}}e^{\alpha\hat{b} \hat{a}^\dagger}\ket{E_0,0}\\
    &=e^{\frac{\abs{\alpha}^2}{2}}\bra{E_{-k},n} \sum_{m,\ell=0}^\infty \frac{\round{-\alpha^*}^m\round{\hat{b}^\dagger}^m \hat{a}^m}{m!}
    \frac{\alpha^\ell \hat{b}^\ell \round{\hat{a}^\dagger}^\ell }{\ell!} \ket{E_0,0}\\
    &=e^{\frac{\abs{\alpha}^2}{2}}\sum_{m,\ell=0}^\infty
    \bra{E_{-k-m},n+m} \frac{\sqrt{\round{n+m}!}}{\sqrt{n!}}
    \frac{\round{-\alpha^*}^m}{m!}
    \frac{\alpha^\ell}{\ell!} 
    \sqrt{\ell!}\ket{E_{-\ell},\ell}
.\end{align}
Here, I used the Backer-Campbell-Hausdorff formula (BCH) to express the displacement operator(eq. \eqref{BCH_explicitly_for_Displacement_Operator}), and used $\round{a^\dagger}^m\ket{n}=\sqrt{\frac{\round{n+m}!}{n!}} \ket{n+m}$. Using the orthogonality of the electron-energy states and the photon states, one get $\left< E_{-k-m},n+m|E_{-\ell},\ell  \right> = \delta_{-k-m,-\ell}\delta_{n+m,\ell}$, so $n=k$ (conservation of energy) and $\ell=m+n$. So, the above is
\begin{align}
    &=e^{\frac{\abs{\alpha}^2}{2}}\sum_{m=0}^\infty
    \frac{\sqrt{\round{n+m}!}}{\sqrt{n!}}
    \frac{\round{-\alpha^*}^m}{m!}
    \frac{\alpha^m\alpha^n}{\round{n+m}!} 
    \sqrt{\round{n+m}!}\\
    &=e^{\frac{\abs{\alpha}^2}{2}} \frac{\alpha^n}{\sqrt{n!}}
    \sum_{m=0}^\infty
    \frac{\round{-\abs{\alpha}^2}^{m}}{m!}
    =e^{-\frac{\abs{\alpha}^2}{2}} \frac{\alpha^n}{\sqrt{n!}}e^{-\abs{\alpha}^2}\\
    &=e^{-\frac{\abs{\alpha}^2}{2}} \frac{\alpha^n}{\sqrt{n!}}=e^{-\frac{\abs{\alpha}^2}{2}} \frac{\alpha^k}{\sqrt{k!}}\label{state_amplitude_for_electron_cavity_1_st_interaction}
.\end{align}
This is the expected Poisson distribution (see ref. \cite{garcia_de_abajo_multiple_2013})
\begin{equation}
    P_k=\abs{\bra{E_{-k},k} D_{\round{\hat{b}\alpha}}\ket{E_0,0}}^2=e^{-\abs{\alpha}^2} \frac{\alpha^{2k}}{k!}
\end{equation}
Typical EELS is retrieved for $|\alpha|^2 \ll 1$, $Rightarrow P_0\approx \round{1-|\alpha|^2}$, $P_1=|\alpha|^2$.
    
\subsection{PINEM - electron interaction with a strong laser-field}

For PINEM, the initial state $\ket{\psi_i}$, before the electron interacts with light is an uncorrelated state, 
\begin{equation*}
\ket{\psi_i}=\ket{E_0}\otimes \ket{\beta}=\ket{E_0,\beta}
.\end{equation*}
For  large $\beta$ the optical coherent state is a good approximation for classical fields.  The important quantum numbers for the final state are the final quanta of electron-energy gain, $k$, and the remaining number of photons $n$,
\begin{equation*}
\ket{\psi_f^{\text{PINEM}}}=\sum_{n=0}^\infty\sum_{k=-\infty}^\infty c_{n,k}\ket{E_k,n}
.\end{equation*}

 The PINEM interaction can be written by the displacement operator.
\begin{align}
    & c_{n,k}=\bra{E_{k},n} D_{\round{\hat{b}\alpha}}\ket{E_0,\beta} \overset{BCH}{=} \\
    &=e^{\frac{\abs{\alpha}^2}{2}}\bra{E_k,n} e^{-\alpha^*\hat{b}^\dagger \hat{a}}e^{\alpha\hat{b} \hat{a}^\dagger}\ket{E_0,\beta}\\
    &=e^{\frac{\abs{\alpha}^2}{2}}\bra{E_{k},n} \sum_{m,\ell,j=0}^\infty 
    \underbrace{\frac{\round{-\alpha^*}^m\round{\hat{b}^\dagger}^m \hat{a}^m}{m!}}_{e^{-\alpha^*\hat{b}^\dagger \hat{a}}}
    \underbrace{\frac{\alpha^\ell \hat{b}^\ell \round{\hat{a}^\dagger}^\ell }{\ell!}}_{e^{\alpha\hat{b}\hat{a}^\dagger}}
    \underbrace{e^{-\frac{\abs{\beta}^2}{2}}\frac{\beta^j}{\sqrt{j!}}}_{\ket{\beta}} \ket{E_0,j}\\
    &=e^{\frac{\abs{\alpha}^2-\abs{\beta}^2}{2}} \sum_{m,\ell,j=0}^\infty 
    \bra{E_{k-m},n+m}\sqrt{\frac{\round{n+m}!}{n!}}  
    \frac{\round{-\alpha^*}^m}{m!}
    \frac{\alpha^\ell}{\ell!}
    \frac{\beta^j}{\sqrt{j!}} \sqrt{\frac{\round{j+\ell}!}{j!}}\ket{E_{-\ell},j+\ell}
.\end{align}    

\begin{figure}[H]
\centering
\includegraphics[width=0.32\textwidth]{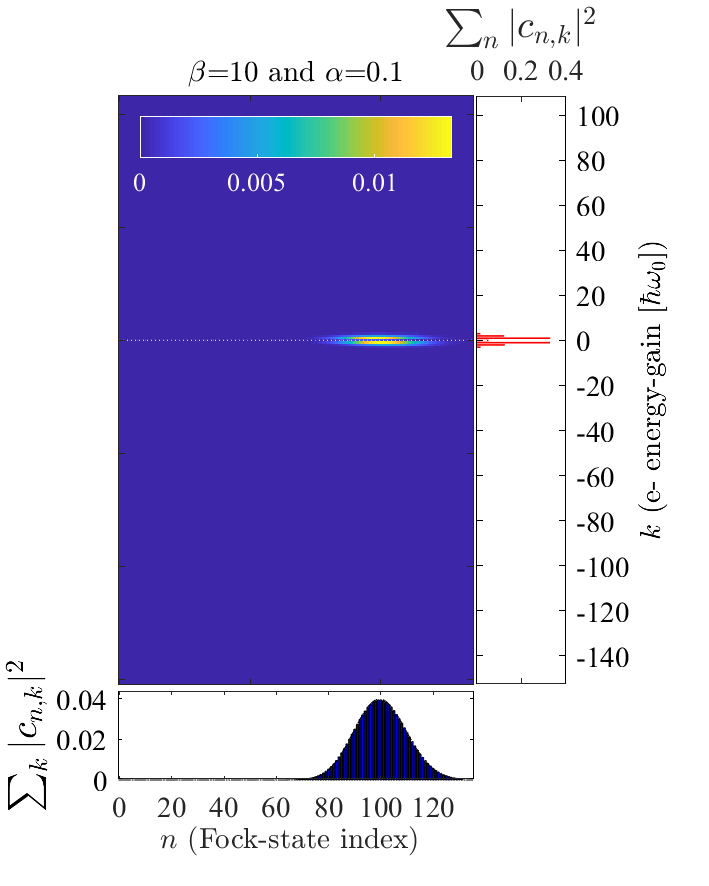}
\includegraphics[width=0.32\textwidth]{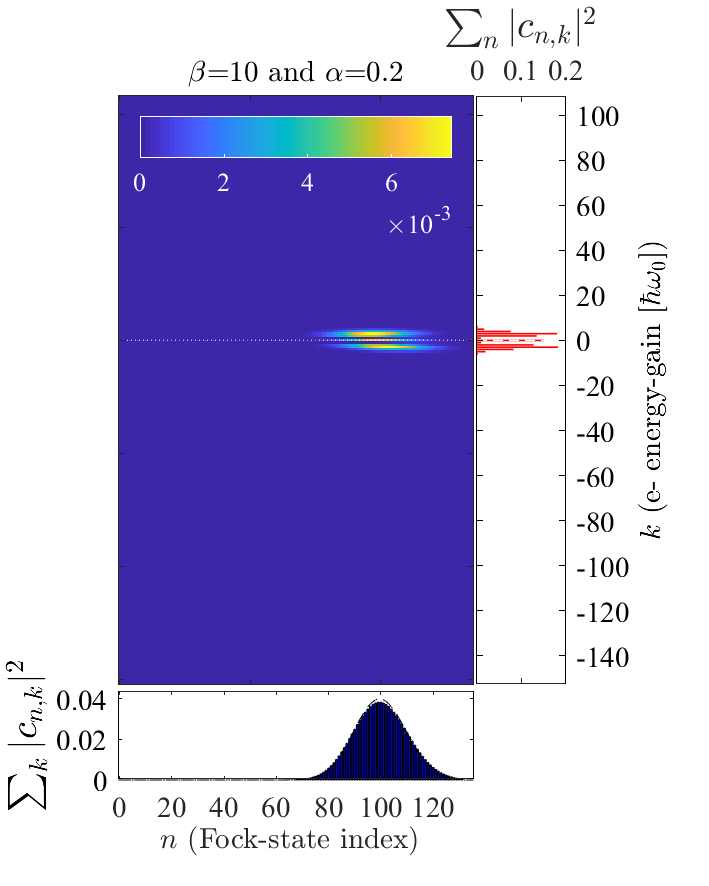}
\includegraphics[width=0.32\textwidth]{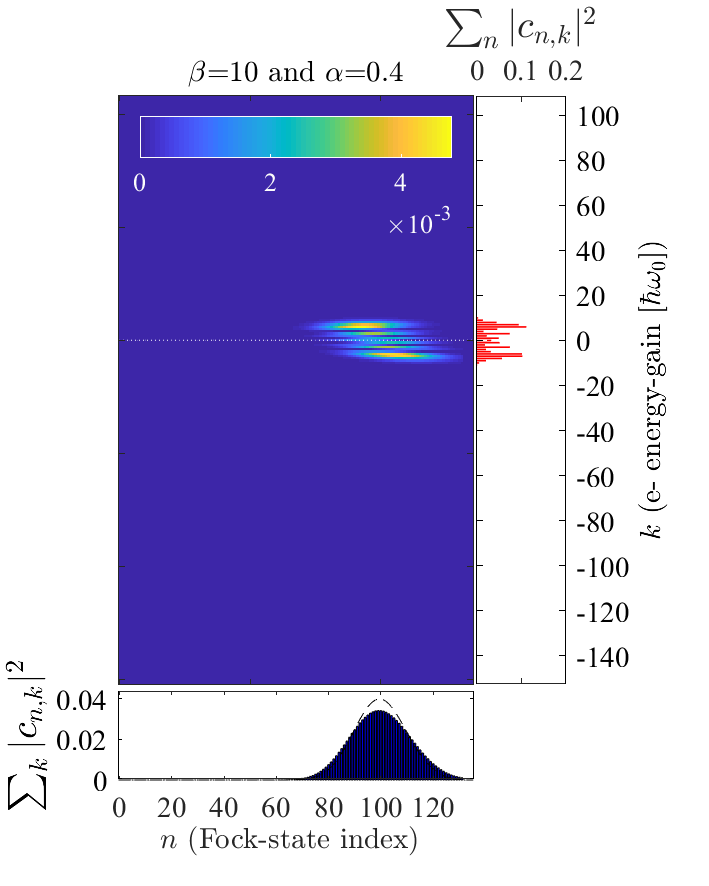}
\includegraphics[width=0.32\textwidth]{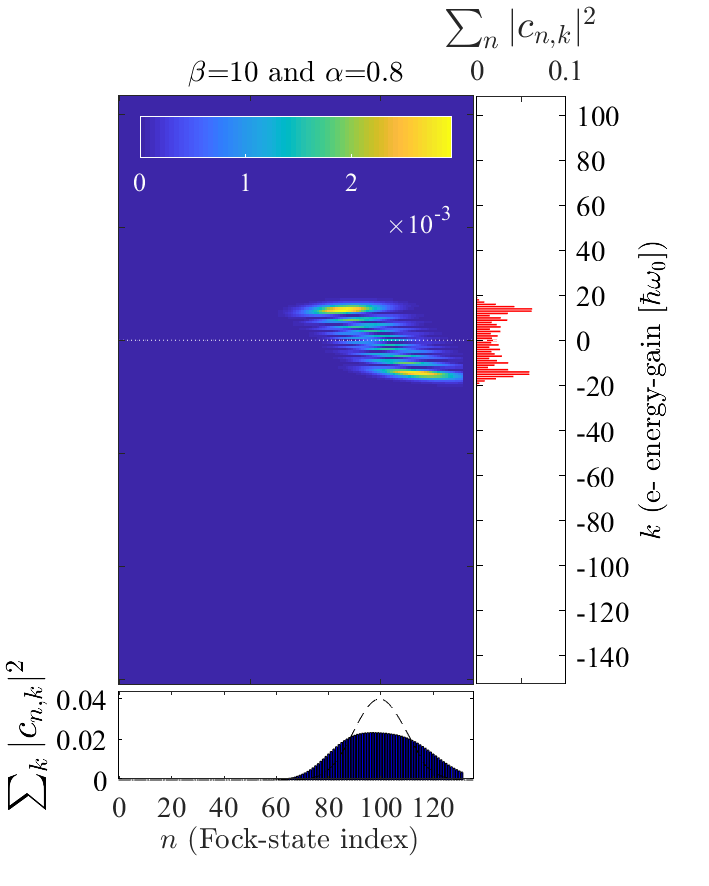}
\includegraphics[width=0.32\textwidth]{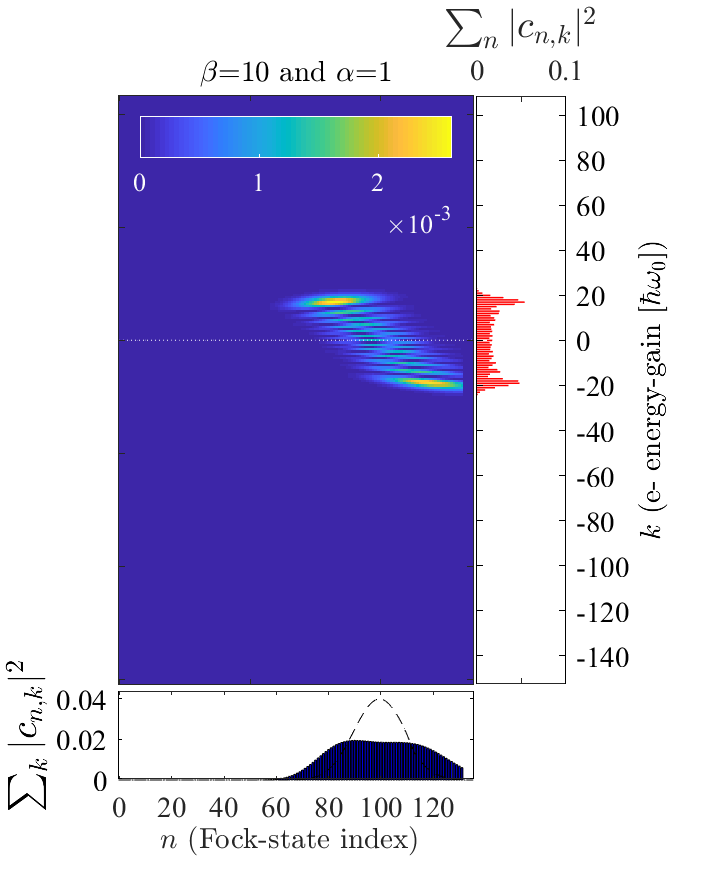}

\caption{The electron-photon spectrogram for various coupling constants, assuming an initial cavity population of 100 photons in the form of a coherent state $\ket{\beta=10}$. The colormap is the co-incidence of a particular energy combination $|c_{n,k}|^2$ of the photon and electron. The bottom axis is the initial (dashed line) and final (blue bars) distribution of the Fock-states for photons in the cavity. The right axis (red bars) is the electron spectrum}
\label{PINEM_like_e_ph_Spectrograms}

\end{figure}

Orthogonality of the states imposes 
\begin{align}
\left<{E_{k-m},n+m}|{E_{-\ell},j+\ell}\right>  &=  \delta_{k-m,-\ell}\delta_{n+m,j+\ell}\\
\text{so, } & m=k+\ell, \, n+m=n+k+\ell=j+\ell.
\end{align}
Thus, one remain with a summation over $\ell$,
\begin{equation}
c_{n,k}=e^{\frac{\abs{\alpha}^2-\abs{\beta}^2}{2}} \sum_{\ell=0}^\infty 
    \sqrt{\frac{\round{n+k+\ell}!}{n!}}  
    \frac{\round{-\alpha^*}^{k+\ell}}{\round{k+\ell}!}
    \frac{\alpha^\ell}{\ell!}
    \frac{\beta^{n+k}}{\round{n+k}!} \sqrt{\round{n+k+\ell}!}
.\end{equation}
After some rearrangements, the final expression for the final-state amplitudes is 
\begin{equation}
   \boxed{c_{n,k} =e^{\frac{\abs{\alpha}^2-\abs{\beta}^2}{2}} \frac{\round{-\alpha^*}^k \beta^{n+k}}{\sqrt{n!}} \sum_{\ell=0}^\infty 
    \underbrace{\frac{\round{n+k+\ell}!}{\round{n+k}!}}_{(*)}  
    \frac{\round{-\abs{\alpha}^2}^{\ell}}{\round{k+\ell}!\ell!}\label{PINEM_expression_k_ge_0}}
.\end{equation}
This is an \textit{exact} expression for the quantum state following PINEM, at any coupling strength. To extend this expression for gain ($k>0$) and for loss ($k<0$) one can replace the factorial operations by Riemann's gamma function $x!\to\Gamma\round{x+1}$, which diverges for negative integers. Since  possible negative factorials terms of $\round{k+\ell}!\ell!$ diverge in the denominator, their corresponding arguments can be ignored. 
For $\round{n+k}<0$, the term marked $\round{*}$, is either $\round{*}=1$ for $\ell=0$, or $\round{*}=0$ for $\ell>0$, and is thus regularized. Section \ref{section_PINEM_Separated_gain_loss} retrieves the explicit PINEM coefficients for gain and loss, using the factorials of explicitly positive integers, rather than Riemann's Gamma-function. One example of PINEM-like spectrum is in the main text (figure 2c), and here, figure\ref{PINEM_like_e_ph_Spectrograms} presents similar spectrograms, for various coupling constants.

\subsubsection{Retrieving the experimental PINEM spectrum for weak interactions with strong fields}
Here, I show how the derivation above retrieves the known PINEM spectrum for the electron, and in what conditions the field is decoupled from the electron state modification. Using eq. \eqref{PINEM_expression_k_ge_0}, in the parameter regime accessible to experiments to date, this derivation should yield the a Bessel-function spectrum - $\propto \abs{J_k\round{2\abs{g}}}^2$, with a possible additional phase. In the experiments the coupling is weak, the field is strong, and there are only few quanta of energy exchanged between the photons and the electrons
\begin{align}
\abs{\alpha}^2 &\ll 1\\
\left< n+k \right>&=\left< j \right>=\abs{\beta}^2\\
\ell  &\lessapprox k \ll n
.\end{align}
Although the summation is up to $\ell\to\infty$, the argument in the sum decays rapidly for $\ell>\abs{\alpha\beta}^2$, so one can compare $\ell$ with other parameters of this system. The comparison to $\abs{\beta\alpha}$ becomes clearer from eq. \eqref{alpha_sqrt_n_k_summation} and the definition of the Bessel-function of the first kind. By employing eq. \eqref{Approximating_N_pule_l_factorial}, and the ratio $\ell/\round{n+k}\ll 1$, one can write $(*)\approx \round{n+k}^{\ell}$.
In that case, the summation arguments acquire the following form
\begin{align}
\frac{\round{n+k+\ell}!}{\round{n+k}!}    \frac{\round{-\abs{\alpha}^2}^{\ell}}{\round{k+\ell}!\ell!}
&\approx\round{n+k}^{\ell}\frac{\round{-\abs{\alpha}^2}^{\ell}}{\round{k+\ell}!\ell!}\\
&=\frac{\round{-\abs{\alpha\sqrt{n+k}}^2}^{\ell}}{\round{k+\ell}!\ell!}\label{alpha_sqrt_n_k_summation}\\
&=\frac{\round{-\abs{g}^2}^{\ell}}{\round{k+\ell}!\ell!}
,\end{align}
 where $g$ is 
 \begin{equation}
 \boxed{g=\alpha\sqrt{n+k}}\approx \alpha\abs{\beta}. \label{g_from_alpha_and_n_photons}
 \end{equation}
This brings the form of the Bessel-function amplitudes to the energy spectrum,
 
\begin{align}
c_{n,k}&=e^{\frac{\overbrace{\abs{\alpha}^2}^{\ll 1}-\abs{\beta}^2}{2}} \frac{\beta^{n+k}}{\sqrt{n!}} e^{ik\arg\round{-\alpha^*}}\round{\frac{2\abs{g}}{2\sqrt{n+k}}}^k \sum_{\ell=0}^\infty   
    \frac{\round{-\frac{\round{2\abs{g}}^2}{4}}^{\ell}}{\round{k+\ell}!\ell!}\\
&=e^{-\frac{\abs{\beta}^2}{2}} \frac{\beta^{n+k}}{\sqrt{n!}} \sqrt{\round{n+k}}^{-k} e^{ik\arg\round{-g^*}} J_k \round{2\abs{g}}
.\end{align}
The above approximation almost reproduces the Bessel-like amplitudes of PINEM, but it leaves some correlations between $n$ and $k$. To remove these correlations and retrieve classical-field effects, one has to neglect correlations in the coherent state. Specifically when assuming $<j>=<n+k>\approx \abs{\beta}^2$ and $\braket{\sqrt{n+k}} \approx \sqrt{\braket{n+k}}$ the following is simplified
\begin{align}
\round{\sqrt{n+k}}^{- k}\approx \round{\beta e^{-i \arg \round{ \beta}}}^{- k} =\beta^{-k}e^{i k\arg \round{ \beta}}\\
\beta^{n+k}\underbrace{\round{\sqrt{n+k}}^{- k}}_{\abs{\beta}^{-k}} & \approx \beta^n \cdot  e^{i k\arg \round{ \beta}}. 
\end{align}
The \textbf{photon states and electron states are now separable}.
\begin{align}
\ket{\psi_f}&=\sum_{n,k} c_{n,k}\ket{E_{k},n}\\
&\approx \left[\sum_n 
e^{\frac{-\abs{\beta}^2}{2}}
 \frac{\beta^n}{\sqrt{n!}} 
 \right] \left[\sum_k
  e^{ik\overbrace{\round{\arg \round{\beta}+\arg\round{-g^*}}}^{=\arg\round{\beta g}  }} J_k \round{2\abs{g}} \ket{E_{k},n} \right]\label{Eq_of_separated_PINEM_1}\\
&= \ket{\beta} \otimes  \sum_k\left[ 
  e^{ik\round{\arg\round{\beta g}}} J_k \round{2\abs{g}} \right] \ket{E_{k}}\\
 &= \ket{\beta} \otimes  \sum_k c_k \ket{E_{k}}
.\end{align}
I used here the relation $g=-g^*$, or $\arg{\round{g}}=\arg{\round{-g^*}}$ since $\alpha=-\alpha^*$.

\subsubsection{Comments on PINEM with nearly classical fields} \label{section_PINEM_classical_fields}
Some points from the above derivation of final state for strong fields interacting weakly with electrons are worth stressing:

\begin{itemize}
\item $g$ has the same meaning as for classical fields, as in Refs. \cite{feist_quantum_2015,echternkamp_ramsey_type_2016}.

\item The electronic states have the amplitude as in the experiments, 
\begin{equation}
c_k=e^{ik\round{\arg\round{\beta g}}} J_k \round{2\abs{g}}
,\end{equation} 
not only the probabilities.

\item $g$ is proportional to the electric field and the coupling constant, $g \propto \alpha \abs{E}$, since $\abs{E}\propto \sqrt{\left< n+k \right>}=\abs{\beta} $. This is in agreement with the it's classical definition. 

\item There is a phase locking between the initial coherent state and the electron state. It appears in the argument $e^{ik\round{\arg\round{\beta g}}}$. 

\item The locking phase just contributes an linear phase with the energy, that is, it provides for the definition of time-zero. 

\item 
$\abs{g}=\abs{\alpha\beta}$, which means that one can increase the width of the electron spectrum (have many PINEM orders, $\Delta E \propto 2\abs{g}$). The scaling, for a given cavity, will be linear with the interaction length (via $\alpha$), and linear with the PINEM-driving electric field (via $\beta$). 

\item 
Since for a coherent state $\ket{\beta}$, the coupling and $g$ are related by $g=\alpha| \beta|$, it means that the coupling, $\alpha$ can be retrieved from classical calculations of PINEM by 
\begin{equation}
\alpha=\frac{g}{\abs{\beta}}\label{alpha_from_g}
.\end{equation}

\item 
As mentioned in the main text, the equivalence of the gain and loss channel originates from a small coupling, $\abs{\alpha} \ll 1$. The mean energy loss is $\abs{\alpha}^2$ in \textit{any} experimental configuration, EELS, PINEM, weak- or strong-coupling. 

\item In practice, the correlations between the photon states $n$ and the electron energy indices $k$ is negligible for a high-$\beta$ coherent state and weak coupling. This is visualy clear from the calculation in the main text.
\end{itemize}

\subsubsection{Separated expressions for gain and for loss channels of electron-photon interactions}
\label{section_PINEM_Separated_gain_loss}
One reason to keep the factorials in eq.\eqref{PINEM_expression_k_ge_0} is the numerical errors induced when evaluating factorials $x!$ through the Riemann Gamma function $\Gamma(x+1)$. Thus, for the calculation presented in the figures of the main text, I explicitly separated the expression to the cases of $k \ge 0$ and $k<0$. This is done by choosing the summation index that spans $0\to \infty$. That is, $\ell$ for $k\ge 0$ and $m$ for $k<0$. The other index is eliminated by the substitution
\begin{align}
\text{for } k \ge 0 &\Rightarrow m=k+\ell \label{PINEM_indices_k_ge_0}\\
\text{for } k < 0 & \Rightarrow  \ell=m-k \label{PINEM_indices_k_le_0}
.\end{align}
Such a separation would explicitly assure that the physical constraint are met $m,\ell,j,n \ge 0$. 
First, for $k \ge 0$ we have eq. \eqref{PINEM_expression_k_ge_0}, with $\ell$ as the summation index. For $k<0$, the index selection in in eq. \eqref{PINEM_indices_k_le_0} is $\ell=m-k$ and $n=j-k$,
\begin{align}
c_{n,k} 
&\overset{\round{k<0,n \ge 0}}{=} e^{\frac{\abs{\alpha}^2 - \abs{\beta}^2}{2}} \frac{\round{-\alpha^*}^k \beta^{n+k}}{\sqrt{n!}} \sum_{m=0}^\infty \frac{\round{n+m}!}{\round{n+k}!}  
    \frac{\round{-\abs{\alpha}^2}^{m-k}}{m!\round{m-k}!}\\
    &= e^{\frac{\abs{\alpha}^2-\abs{\beta}^2}{2}} \frac{\alpha^{-k} \beta^{n+k}}{\round{n+k}!\sqrt{n!}}  
    \sum_{m=0}^\infty \round{n+m}!  
    \frac{\round{-\abs{\alpha}^2}^m}{m!\round{m-k}!}    \\
    &= e^{\frac{\abs{\alpha}^2-\abs{\beta}^2}{2}} \frac{\alpha^{\abs{k}} \beta^{n+k}}{\round{n+k}!\sqrt{n!}}  
    \sum_{m=0}^\infty \round{n+m}!  
    \frac{\round{-\abs{\alpha}^2}^m}{m!\round{m+\abs{k}}!} \label{PINEM_expression_k_le_0}
\end{align}
The last expressions are retrieved by changing t $-k \to \abs{k}$, and using $\alpha=-\alpha^*$. Thus, the two expressions for gain and loss,  eqs. \eqref{PINEM_expression_k_ge_0} and  \eqref{PINEM_expression_k_le_0}, respectively, can be combined  
\begin{align}
\ket{\psi_f^{PINEM}}&=\sum_{n=0}^\infty \sum_{k=-\infty}^\infty c_{n,k} \ket{E_k,n}\\
c_{n,k} 
&=e^{\frac{\abs{\alpha}^2 - \abs{\beta}^2}{2}} \frac{\alpha^{\abs{k}}\beta^{n+k}}{\round{n+k}!\sqrt{n!}}  
 \sum_{\ell=0}^\infty 
       \frac{\round{-\abs{\alpha}^2}^\ell}{\round{\ell+\abs{k}}!\ell!}  
  \cdot \left\lbrace
  \begin{array}{lr}
 \round{n+k+\ell}! & \text{ for } k \ge 0\\
\round{n+\ell}! & \text{ for } k<0
 \end{array}\right.\label{PINEM_c_n_k_separated_cases}
.\end{align}
Note that for the $k<0$ part, I just wrote the arbitrary summation index as $\ell$ instead of $m$, and used $-k=\abs{k}$, while for the $k>0$ part, I substituted $k=\abs{k}$.

\section{Two-electron interaction mediated by cavity photons}
After an interaction of an electron with a cavity, eq. \eqref{state_amplitude_for_electron_cavity_1_st_interaction} calculates the final state amplitudes,
\begin{align}
\ket{\psi}=\sum_{\kappn} e^{-\frac{\abs{\alpha_1}^2}{2}} \frac{\alpha_1^{\kappn}}{\sqrt{\kappn!}} \ket{E_{-\kappn},\kappn}
.\end{align}
It is a coherent photonic state with parameter $\alpha_1$, the strength of the first interaction, and an electronic part that conserves a net energy $E_0$. I derive the interaction strength as $\alpha_1$ and $\alpha_2$ for the first and second electron, respectively, to be able to separate their contributions. Typically, one can take equal interaction strengths, $\alpha_1=\alpha_2$, as done in the main text. The loss index is changed here to ($\kappn$) to differ the loss of the first electron from the gain index of the second electron, $k$ and the photons' index $n$. In the manuscript the index $n$ instead of $s$ for brevity.
I now consider a second electron with energy $\E_0=E_0$. The different symbol just marks a difference between the first and the second electrons. Thus, the initial electron-electron-photon state, before the second electron interacts with stored photons is
\begin{align}
\ket{\psi_i}=\sum_{\kappn} e^{-\frac{\abs{\alpha_1}^2}{2}} \frac{\alpha_1^\kappn}{\sqrt{\kappn!}} \ket{E_{-\kappn},\E_0,\kappn}
.\end{align}
The final state can be characterized by the individual electron-electron-photon states,
\[
\ket{\psi_f^{e-e}}=\sum_{\kappn \ge 0,\, k}c_{\kappn,k} \ket{E_{-\kappn},\E_k,n}
.\]
The coefficients are given by the projection 
\begin{align}
    c_{\kappn,k} &=\bra{E_{-\kappn},\E_k,n} D_{\round{\hat{b}\alpha_2}}\ket{\psi_i}\\
     &=\bra{E_{-\kappn},\E_k,n} \sum_{\kappn,j} e^{-\frac{\abs{\alpha_1}^2}{2}} \frac{\alpha_1^\kappn}{\sqrt{\kappn!}} D_{\round{\hat{b}\alpha_2}} \ket{E_{-\kappn},\E_0,\kappn}\\
     &=\bra{E_{-\kappn},\E_k,n} \sum_{\kappn,j} e^{-\frac{\abs{\alpha_1}^2}{2}} \frac{\alpha_1^\kappn}{\sqrt{\kappn!}} e^{\frac{\abs{\alpha_2}^2}{2}} \sum_{m,\ell}\underbrace{\frac{\round{-\alpha_2^*}^m\round{\hat{b}^\dagger}^m \hat{a}^m}{m!}}_{e^{-\alpha_2^*\hat{b}^\dagger \hat{a}}}
    \underbrace{\frac{\alpha_2^\ell \hat{b}^\ell \round{\hat{a}^\dagger}^\ell }{\ell!}}_{e^{\alpha_2\hat{b}\hat{a}^\dagger}}
 \ket{E_{-\kappn},\E_0,\kappn}\\
      &= \sum_{m,\ell}
      \bra{E_{-\kappn},\E_{k-m},n+m} \sum_{\kappn,j} e^{-\frac{\abs{\alpha_1}^2}{2}} \frac{\alpha_1^\kappn}{\sqrt{\kappn!}} e^{\frac{\abs{\alpha_2}^2}{2}} \sqrt{\frac{\round{n+m}!}{n!}}
      \frac{\round{-\alpha_2^*}^m}{m!}\cdot \label{electron_intermediate_step_1}
\\ 
&\quad\quad\quad\quad\quad\quad\quad\quad
 \frac{\alpha_2^\ell}{\ell!}\sqrt{\frac{\round{\kappn+\ell}!}{\kappn!}}
 \ket{E_{-\kappn},\E_{-\ell},\kappn+\ell}
.\end{align}
Note that here the operator $\hat{b}$ is acting on the second electron, leaving the first unchanged. The states' orthogonality imposes 
\begin{align}
\left< {E_{-\kappn},\E_{k-m},n+m} |{E_{-\kappn},\E_{-\ell},\kappn+\ell} \right>&=\delta_{n+m,\kappn+\ell}\delta_{m-k,\ell} \\
 \Rightarrow m&=\ell+k \\
  \kappn&=n+k.
\end{align} 
Inserting the indices selection, and using the states' orthogonality, eq. \eqref{electron_intermediate_step_1} is 
\begin{align}
      &= \sum_{\ell=0}^\infty e^{-\frac{\abs{\alpha_1}^2}{2}} \frac{\alpha_1^\kappn}{\sqrt{\kappn!}} e^{\frac{\abs{\alpha_2}^2}{2}} \sqrt{\frac{\round{n+\ell+k}!}{n!}}
      \frac{\round{-\alpha_2^*}^\ell\round{-\alpha_2^*}^k}{\round{\ell+k}!}
 \frac{\alpha_2^{\ell}}{\ell!}\sqrt{\frac{\round{n+\ell+k}!}{\round{n+k}!}}\\
 &=  e^{-\frac{\abs{\alpha_1}^2}{2}} \frac{\alpha_1^\kappn}{\sqrt{\kappn!}} e^{\frac{\abs{\alpha_2}^2}{2}} \frac{\round{-\alpha_2^*}^k}{\sqrt{n!\round{n+k}!}}
 \sum_{\ell=0}^\infty \round{n+\ell+k}!
       \frac{\round{-\abs{\alpha_2}^2}^\ell}{\round{\ell+k}!\ell!}\\
.\end{align}
The relation $n+k=\kappn$ allows the last equation to be written with as a function of the electron energies only, $k,  \kappn$. So, one can write the coefficients $c_{\kappn,k}$    
\begin{equation}
\boxed{   
c_{\kappn,k>0}=  e^{-\frac{\abs{\alpha_1}^2}{2}} \alpha_1^\kappn e^{\frac{\abs{\alpha_2}^2}{2}} \frac{\round{-\alpha_2^*}^k}{\sqrt{\round{\kappn-k}!}}
 \sum_{\ell=0}^\infty \frac{\round{\ell+\kappn}!}{\kappn!} \frac{\round{-\abs{\alpha_2}^2}^\ell}{\round{\ell+k}! \ell!}
 }\label{2_electron_EELS_amplitues}
.\end{equation}    
Factorials of negative numbers diverge according to the Riemann's Gamma function. The term $\sqrt{\round{\kappn-k}!}\geq 0$ diverges for $s<k$. In other words, the highest k is the full conversion energy taken from the $1^{st}$ electron to the $2^{nd}$. Thus, this term nullifies the probability for a non-physical energy-gain of the second electron.
Note that in the main text I chose to simplify the system by choosing $\alpha_1=\alpha_2=\alpha$, which is a realistic realization of eq. \eqref{2_electron_EELS_amplitues}, when the two electrons share the same beam path and interact with the same cavity.

Similar to the discussion in section \ref{section_PINEM_Separated_gain_loss}, the indices selection $m,\ell$ differs for the gain- and loss-channels, as used in practice to calculate the 2-particle amplitudes numerically, with the substitutions
\begin{align}
m=\ell+k & \text{ for }k \ge 0\\
\ell= m-k & \text{ for }k<0 \label{indices_m_ell_k_1}
.\end{align} 
First, \textbf{for the case of energy gain by the second electron}, $k>0$, the two-electron probabilty amplitudes are in eq. \eqref{2_electron_EELS_amplitues}.
For the case of \textbf{energy loss by the second electron}, $k<0$, the proper index to keep is $m$, with the relation $\ell=m-k=m+\ktl$ from eq. \eqref{indices_m_ell_k_1}. 
\begin{align}
      &= \sum_{m=0}^\infty e^{-\frac{\abs{\alpha_1}^2}{2}} \frac{\alpha_1^\kappn}{\sqrt{\kappn!}} e^{\frac{\abs{\alpha_2}^2}{2}} \sqrt{\frac{\round{n+m}!}{n!}} \frac{\round{-\alpha_2^*}^{m}}{m!}
 \frac{\alpha_2^{m}\alpha_2^{\ktl}}{\round{m+\ktl}!}\sqrt{\frac{\round{n+m}!}{\round{\kappn}!}}\\
 &=  e^{-\frac{\abs{\alpha_1}^2}{2}} \alpha_1^\kappn e^{\frac{\abs{\alpha_2}^2}{2}} \frac{\alpha_2^\ktl}{\sqrt{\round{\kappn-k}!}} 
 \sum_{m=0}^\infty \frac{\round{m+\round{\kappn+\ktl}}!}{\kappn!}\frac{\round{-\abs{\alpha_2}^2}^m}{\round{m+\ktl}!m!}
.\end{align}
Finally, one can combine the expressions for the energy-gain and energy-loss for the second electron,
 \begin{align}
\ket{\psi_f^{e-e}}&=\sum_\kappn \sum_{k \le \kappn} c_{\kappn,k} \ket{E_{-\kappn},\E_k,\kappn-k}\\
c_{\kappn,k}&=e^{-\frac{\abs{\alpha_1}^2}{2}}  e^{\frac{\abs{\alpha_2}^2}{2}} \frac{ \alpha_1^\kappn  \alpha_2^{\ktl}}{\sqrt{\round{\kappn-k}!}} 
  \left\lbrace
  \begin{array}{lr}
 \sum_{\ell=0}^\infty \frac{\round{\ell+\kappn}!}{\kappn!}
       \frac{\round{-\abs{\alpha_2}^2}^\ell}{\round{\ell+k}!\ell!} & \text{ for }\kappn \ge k \ge 0\\
\sum_{m=0}^\infty \frac{\round{m+\round{\kappn+\ktl}}!}{\kappn!}\frac{\round{-\abs{\alpha_2}^2}^m}{\round{m+\ktl}!m!} & \text{ for\quad\quad\,} k<0
 \end{array}\right.
 \end{align}
Using the equality $-\alpha_2^*=\alpha_2$, setting $\abs{k}$ appropriately, and using just either $\ell$ as a summation index, a more compact equation can be written 
 \begin{equation}
  {c_{\kappn,k}=e^{-\frac{\abs{\alpha_1}^2}{2}} \frac{\alpha_1^\kappn}{\kappn!} e^{\frac{\abs{\alpha_2}^2}{2}} \frac{\alpha_2^{\abs{k}}}{\sqrt{\round{\kappn-k}!}} 
\sum_{\ell=0}^\infty 
       \frac{\round{-\abs{\alpha_2}^2}^\ell}{\round{\ell+\abs{k}}!\ell!}  
  \cdot \left\lbrace
  \begin{array}{lr}
 \round{\ell+\kappn}! & \text{ for }\kappn \ge k \ge 0\\
\round{\ell+\kappn+\abs{k}}! & \text{ for\quad\quad\,} k<0
 \end{array}\right.}\label{c_kappa_k_2nd_electron}
.\end{equation}  
 Except for the spectrograms in the main text, for the coefficients $c_{\kappn,k}$ I added some here, for different coupling strength, $\alpha=\alpha_1=\alpha_2$

\begin{figure}[H]
\centering
\includegraphics[width=0.32\textwidth]{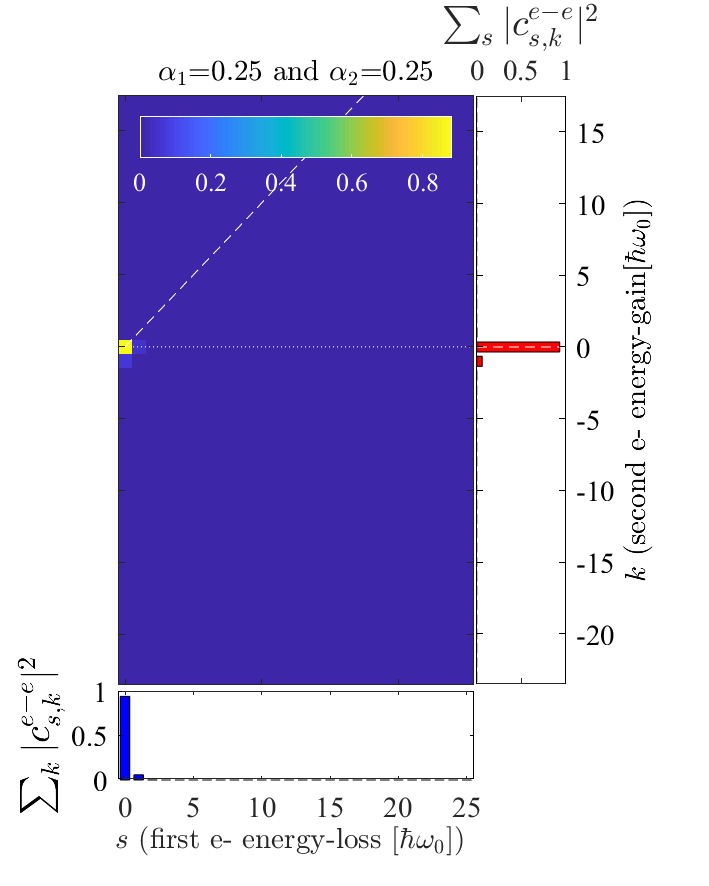}
\includegraphics[width=0.32\textwidth]{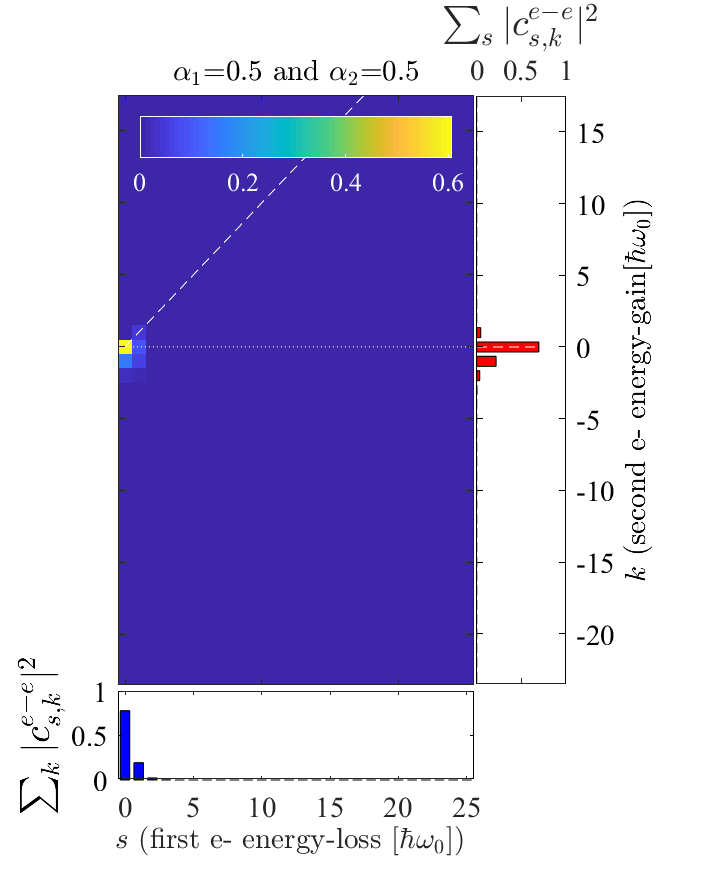}
\includegraphics[width=0.32\textwidth]{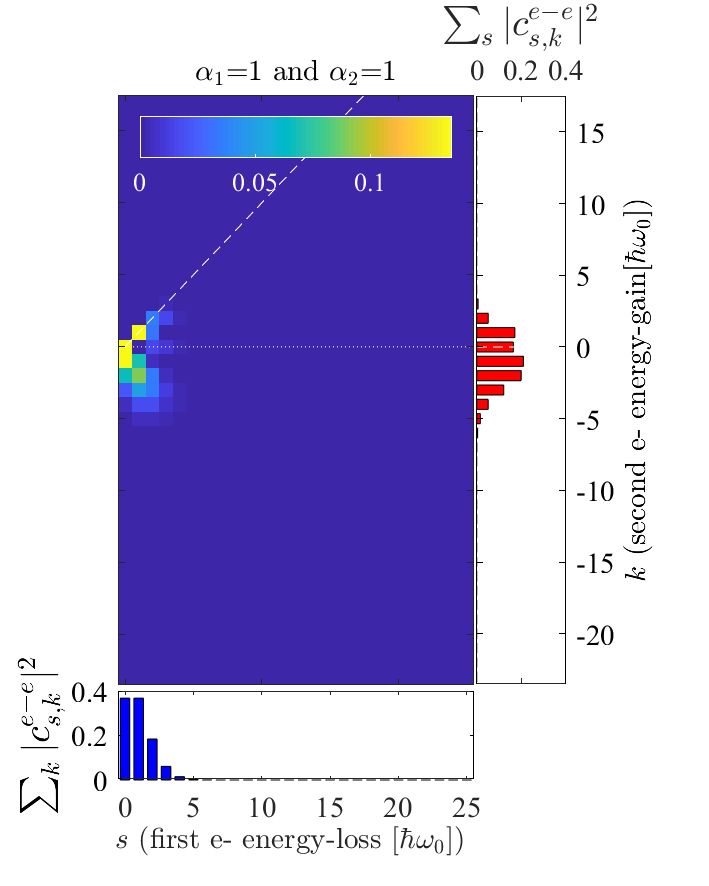}
\includegraphics[width=0.32\textwidth]{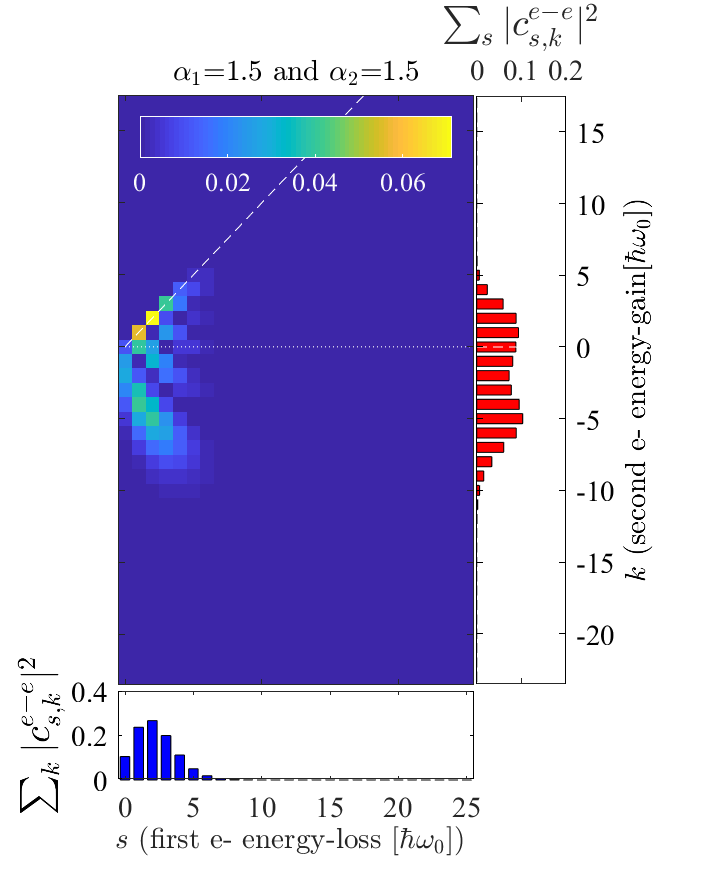}
\includegraphics[width=0.32\textwidth]{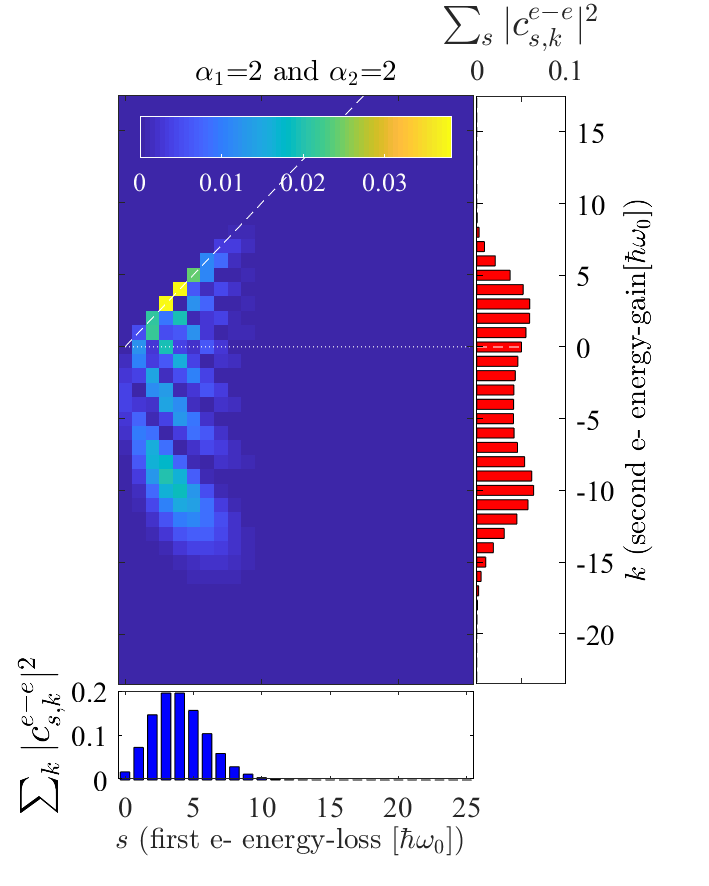}
\includegraphics[width=0.32\textwidth]{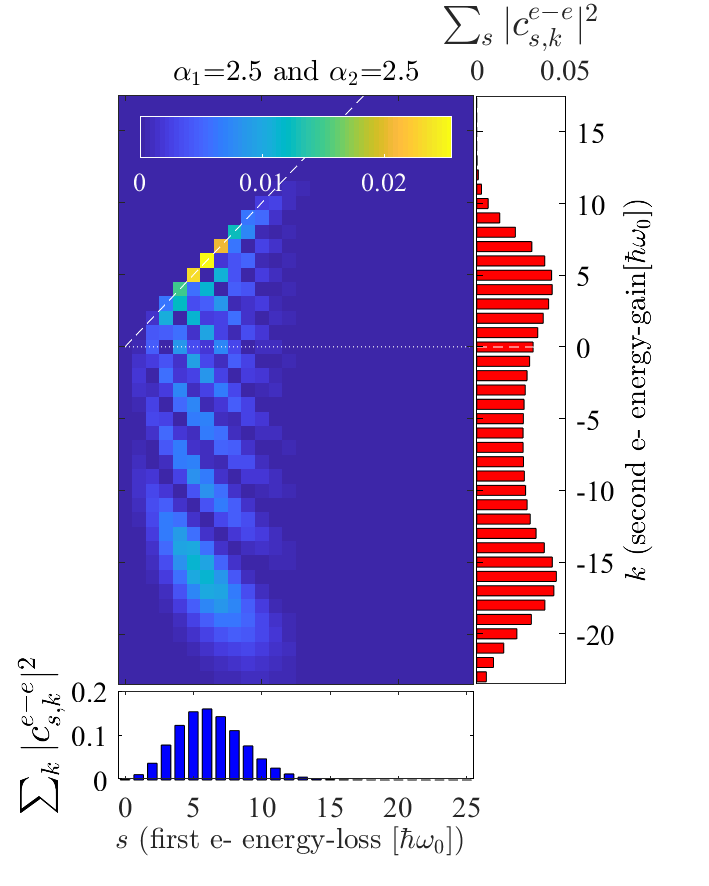}

\caption{The two-electron spectrogram for various values of equal coupling constants. The colormap is the co-incidence of a particular energy comibination $|c_{s,k}|^2$. The bottom axis (blue bars) is the spectrum of the first electron (including only loss channels) and the right axis (red bars) is the spectrum of the second electorn. Note that in the text, the loss index of the first electron is $n$ for brevity, while here it is $\kappn$.}
\label{e_e_Spectrograms}

\end{figure}

There are several interesting examples to consider
\begin{enumerate}

\item
Checking the limit of no initial gain -- if the interaction strength of the first electron is nullified, $\abs{\alpha_1}=0$, the interaction of the second electron should result in the spectrum of a single interaction. Only the coefficients with $\kappn=0$ survive due to the factor $\alpha_1^\kappn$, and the $\sqrt{\kappn-k}$ term suggests that $k \le 0$. The second electron therefore, populates no gain states. According to eq. \eqref{c_kappa_k_2nd_electron}, the coefficients $c_{\kappn,k}$ will be
\begin{align}
  c_{0,k<0} &=  e^{\frac{\abs{\alpha_2}^2}{2}} \frac{1}{\sqrt{\round{-k}!}} 
\alpha_2^{-k}\sum_{m=0}^\infty \frac{\round{m+\round{-k}}!}{\round{m-k}!}\frac{\round{-\abs{\alpha_2}^2}^m}{m!}\\
&=  e^{\frac{-\abs{\alpha_2}^2}{2}} \frac{\alpha_2^{\abs{k}}}{\sqrt{\abs{k}!}} 
,\end{align}  
which is a Poissonian probability distribution, just as in eq. \eqref{state_amplitude_for_electron_cavity_1_st_interaction} .

\item The coefficients retrieved from the two-electron interaction (eq. \eqref{c_kappa_k_2nd_electron}), and for the PINEM interaction (eq. \eqref{PINEM_c_n_k_separated_cases}) are equivalent, by just selecting indices. Since the first electron in the two-electron case induces a coherent state, the following stage, which is the interaction of a coherent state with an approaching electron is identical to strong-coupling PINEM.  However, the important difference is the quantum numbers. For PINEM, the quantum numbers (in which one may search for entanglements) are $n,k$, while for two electrons, their energy states, $\kappn,k$ are important. Thus, for physically relevant purposes, the two cases are sheared $n=\kappn-k$. Such a shear can be identified by comparing figure 2 and figure 3 in the main text. One can also consider the $1^{st}$ electron spectrum as corresponding to the initial optical state, which differs from the final, non-coherent-state-like photon distribution.

\item 
\textbf{Approaching strong-field PINEM for the $2^{nd}$ electron}.\\
Similar to section \ref{section_PINEM_classical_fields}, $c_{\kappn,k}$ can resemble the experimentally measured PINEM for weak coupling limit. For the two-electron case, that requires verri different coupling strengths, $\abs{\alpha_1 \gg 1 \gg \abs{\alpha_2}}$.For a large energy deposition in the cavity by the first electron, one can assume  $\round{\ell+\abs{k}}, \ell \ll \kappn $, and approximate
\begin{align}
\round{\ell+\kappn+\abs{k}}! & \approx \kappn!\kappn^\ell \kappn^{\abs{k}}\\
\Rightarrow \round{\ell+\kappn+\abs{k}}!\round{-\abs{\alpha_2}^2}^\ell &=\kappn! \kappn^{\abs{k}}\round{-\abs{\alpha_2\sqrt{\kappn}}^2}^\ell
.\end{align}
The factorial approximation is justified in eq. \eqref{Approximating_N_pule_l_factorial}. Considering the gain/loss dependent part of \eqref{c_kappa_k_2nd_electron}, including the term $\sqrt{\round{\kappn-k}!}$, and changing $k$ to $\abs{k}$ in a consistent manner for $k<0$ and $k\ge0$, one can write the following equalities to combine the gain and loss parts
\begin{align}
 \left\lbrace
  \begin{array}{lr}
 \round{\ell+\kappn}!\round{\round{\kappn-\abs{k}}!}^{-\half} \\
\round{\ell+\kappn+\abs{k}}!\round{\round{\kappn+\abs{k}}!}^{-\half} 
 \end{array}\right.
&=  \left\lbrace
  \begin{array}{lr}
 \kappn!\kappn^\ell \round{\kappn!}^{-\half}\kappn^{\frac{\abs{k}}{2}} & \text{ for }\kappn \ge k \ge 0\\
\kappn!\kappn^\ell\kappn^{\abs{k}}\round{\kappn!}^{-\half} \kappn^{-\frac{\abs{k}}{2}} & \text{ for\quad\quad\,} k<0
 \end{array}\right.\\
 &= \sqrt{\kappn!} \round{\sqrt{\kappn}}^{\abs{k}} \kappn^\ell
,\end{align}  
Incorporating that into eq. \eqref{c_kappa_k_2nd_electron} gives a separable equation. Same logic here applied as 
\begin{align}
c_{\kappn,k}&=\left[e^{-\frac{\abs{\alpha_1}^2}{2}} \frac{\alpha_1^\kappn}{\sqrt{\kappn!}}\right] 
\left[ e^{\frac{\abs{\alpha_2}^2}{2}} \underbrace{\round{\alpha_2\sqrt{\kappn}}^{\abs{k}} 
\sum_{\ell=0}^\infty 
       \frac{\round{-\abs{\alpha_2 \sqrt{\kappn}}^2}^\ell}{\round{\ell+\abs{k}}!\ell!}}_{J_{k}\round{2\abs{g}}}\right]
,\end{align}
where, again, $g=\alpha_2\sqrt{\kappn}\approx \alpha_2\abs{\alpha_1}$, and this retrieves the experimental Bessel-amplitudes of PINEM.  One should note that the above separability is naturally occurring  for the quantum numbers $\kappn,k$ since the first electron is unaffected by any detail of the intaraction with a second elctron. It already left the interaction region. In the PINEM case, where the final state is expressed with $n,k$ and $\kappn=n+k$, the two states cannot be separated. A similar treatment leading to eq. \eqref{Eq_of_separated_PINEM_1} would result in $c_{n,k}=\left[ e^{-\frac{\abs{\beta}^2}{2}}\frac{\beta^{n+k}}{\sqrt{n+k}} \right]\left[ \cdots\right]$, which is clearly not separable. For this reason, the quantum fluctuations remain here, while they have to be mitigated in eq. \eqref{Eq_of_separated_PINEM_1}. 
\end{enumerate}

\section{Ladder operators for the relativistic electron }
\label{section_for_rel_electrons} 

\begin{itemize}
        \item 
        Assuming the energy is allowed in levels $\ket{n}$, the Hamiltonian comply with $\hat{H}\ket{n}=E_n\ket{n}$. 
        \item 
        In this section, $n$ is the energy level-index of the electron, with respect to the zero-loss energy $E_{n=0}=E_0$. The number operator is \[\hat{n}=\frac{\hat{H}-E_0}{\hbar\omega_0}.\] 
        For the zero-loss energy, $\hat{H}\ket{0}=E_0\ket{0}$. $E_0$ relates to the electron rest energy $E_{rest}$ and the zero-loss momentum $P_0$ by  $E_{n=0}=\sqrt{E_{rest}^2+\round{P_0c}^2}$. 
        
    \item 
    For nearly plane-wave electrons, the ladder operators commute $[b^\dagger,b]=0$. In that case, They cannot construct the Hamiltonian. The Hamiltonian is, by definition, sensitive to the level index, and hence cannot commute with a ladder operator, e.g. $\hat{H}\round{b^\dagger\ket{n}}=E_{n+1}\round{b^\dagger\ket{n}}\neq b^\dagger\round{\hat{H}\ket{n}}=E_n b^\dagger\round{\ket{n}}$.
    \item The momentum of state $\ket{n}$ is $P_0+P_n$. Thus, it can be written as $\ket{n}=\exp\left[{\frac{i}{\hbar}\round{P_0+P_n}}\right]$
    
    \item
    The dispersion relation for the electrons around the zero-loss energy is
    \begin{align}
    E=&\sqrt{E_{rest}^2+\round{P_0+P_n}^2c^2}\\
    =&\sqrt{E_{rest}^2+\round{P_0c}^2}\sqrt{1+\frac{2 P_0 P_nc^2+P_n^2c^2}{E_{rest}^2+\round{P_0c}^2}}\\
    \approx &\sqrt{E_{rest}^2+\round{P_0c}^2}\round{1+\half\frac{2 P_0 P_n c^2+P_n^2c^2}{E_{rest}^2+\round{P_0c}^2}}\\
    =&const 
    +\frac{P_0 P_n c^2}{\sqrt{E_{rest}^2+\round{P_0c}^2}}
    +\frac{P_n^2c^2}{2\sqrt{E_{rest}^2+\round{P_0c}^2}}\\
    =&E_{Zero-loss} 
    +v_{Zero-loss} \cdot P_n \round{1+\half\frac{P_n}{P_0}} \label{H_electrons_E_P_P2}
    \end{align}
        \item 
        For relativistic electrons, $P_n \ll P_0$, the Hamiltonian is linear with the momentum, $P_n$. The ladder operators can be written explicitly as \[\hat{b}^\dagger=e^{i\Delta k x}\text{ and } \hat{b}=e^{-i\Delta k x},\] with $\Delta k=\round{P_{n+1}-P_n}/\hbar=\omega_0/{v_{Zero-loss}}$. 

    \item 
    To show the ladder operators are correct, one needs to show is that $\hat{H}\round{b^\dagger\ket{n}}=E_{n+1}\round{b^\dagger\ket{n}}$
    \begin{align}
        \hat{H}\round{b^\dagger\ket{n}}=&\hat{H} \round{e^{i\frac{\omega_0}{c}x} \cdot e^{i(P_0+P_n)x}}\\
        =&\hat{H} \cdot e^{i(P_0+P_n+\frac{\omega_0}{c})x}\\
        =&\hat{H} \cdot e^{i(P_0+P_{n+1})x}\\
        =&E_{n+1} e^{i(P_0+P_{n+1})x}\\
        =&E_{n+1} \round{e^{i\frac{\omega_0}{c}x} \cdot e^{i(P_0+P_n)x}}
        =E_{n+1}\round{b^\dagger\ket{n}}.
    \end{align}
    Similarly, $\hat{H}\round{b\ket{n}}=E_{n-1}\round{b\ket{n}}$. 
	\item 
	Since $\hat{b},\hat{b}^\dagger$ are pure phasors, they reconstruct the relations in Ref. \cite{feist_quantum_2015},
\begin{align}
        b^\dagger\ket{n}=&\ket{n+1}\\
        b\ket{n}=&\ket{n-1}.
    \end{align}
This relation applies also for non-relativistic electrons if the underlying assumptions hold.
	\item 
	Alternatively, the ladder operators can be constructed in a diagonal form
\begin{align}
\hat{b}&=\sum_n e^{i\round{k_{n-1}-k_n}x}\ket{n}\bra{n}\\
\hat{b}^\dagger&=\sum_n e^{i\round{k_{n+1}-k_n}x}\ket{n}\bra{n}.
,\end{align} 
based on the known values of the wave-vectors $k_n$. One can see that, especially for slow electrons $\hat{b}$ and $\hat{b}^\dagger$ are not exactly hermitian conjugates of each other. Hence, they may not be convenient for the representation of observables quantities. However, even at acceleration voltages of few keV, the dispersion becomes linear enough to allow the assumption that $\hat{b}$ and $\hat{b}^\dagger$ are complex conjugates. 
\end{itemize}

\section{Quantitative evaluation of electron-fiber coupling \label{section_quantization_of_fiber_modes} } 
This section is mostly technical, in the form of bullet-points that allows, with the sources to follow the quantitative estimation of the coupling constant (say, per $1\mu m$). It is based on Refs. \cite{yeh_guided_wave_1987,chen_foundations_2006} and assisted by Prof. Elias N. Glytsis notes about Cylindrical Dielectric Waveguides, 2017. Numerical results from finite-elements calculation are in good agreement with the analytic calculation below.  (see figure \ref{COMSOL_463_nm_Si3N4})

 In short, I calculate the field distribution for an $HE_{11}$ mode in a round , clad-less, step-index fiber, and estimate the field close to it's surface, in vacuum. For a given fiber length of $1\,\mu m$, I calculate $g$ and employ eq. \eqref{alpha_from_g}, to calculate $g$ per photon, that is, the coupling constant $\alpha$. The average number of photons in the classical field of the mode is evaluated as $<n>=U/\hbar\omega$, where $U$ is the total field's energy, per $\mu m$, and $\hbar\omega$ is the photon energy. \\
I start with the basic form of the mode. An $HE_{11}$ mode is always guided in a fiber. It is typically given by the the electric and magnetic fields parallel to the fiber , $E_z$ and $H_z$
\begin{align}
E_z^{HE_{11}}(r,\phi,z,t) &=e^{i\omega t-i\beta z}{\sin\phi}
\begin{cases}
A_1J_1\round{u\frac{r}{a}} & r \le a \\
B_1K_1\round{w\frac{r}{a}} & r > a \\
\end{cases}\\
H_z^{HE_{11}}(r,\phi,z,t) &=e^{i\omega t-i\beta z}{\sin\phi}
\begin{cases}
F_1J_1\round{u\frac{r}{a}} & r \le a \\
G_1K_1\round{w\frac{r}{a}} & r > a \\
\end{cases}\label{H_z_round_fiber}\\
u=\sqrt{k_{in}^2-\beta^2}\quad , &\quad k_{in}=\frac{2\pi}{\lambda}n_{core}\\
w=\sqrt{\beta^2-k_{out}^2}\quad , &\quad k_{out}=\frac{2\pi}{\lambda}n_{clad=vacuum}
.\end{align}
$J_\ell(x)$, and $K_\ell(x)$ are the Bessel function of the first kind and the modified Bessel function of the second kind. An electron traversing parallel to the fiber will excite one linearly polarized mode, thus, radial function is $\sin\round{\phi}$. The normalization for the radial function is already included in the calculation of $A_1$.\\
First, I note that $E_z$ is the most relevant field component, as it determines $g$ by accelerating or decelerating the co-propagating electrons.

 For $HE_{11}$, find the smallest solution of the propatation constant, $\beta$, from the equation for $\ell=1$. 
\begin{align}
\left[ \frac{1}{u}\frac{J'_\ell(u)}{J_\ell(u)} + \frac{1}{w}\frac{K'_\ell(w)}{K_\ell(w)}  \right]
\left[ \round{\frac{n_{core}}{n_{clad}}}^2\frac{1}{u}\frac{J'_\ell(u)}{J_\ell(u)} + \frac{1}{w}\frac{K'_\ell(w)}{K_\ell(w)}  \right]
&=\round{\frac{\beta\ell}{k_{out}}}^2\left[ \frac{1}{u^2}+\frac{1}{w^2}  \right]^2
.\end{align}
Here, $J'_\ell(u)=\left.\frac{d}{dx}J_\ell(x)\right|_{x=u}$, and similarly for $K'_\ell(w)$. From $\beta$, one finds $u$ and $w$.
 The remaining coefficients, $B_1,F_1,G_1$ (assuming an arbitrary $A_1=1$ for simplicity) are, 
\begin{align}
A_1&=1 \label{Amplitudes_A1_eq_1}\\
B_1&=\frac{J_\ell(u)}{K_\ell(w)} A_1\\
G_1&=\frac{J_\ell(u)}{K_\ell(w)} F_1\\
F_1&=\frac{1}{\mu_0 \omega}\round{i\beta\ell}\round{\frac{1}{u^2}+\frac{1}{w^2}}\left[ \frac{1}{u}\frac{J'_\ell(u)}{J_\ell(u)} + \frac{1}{w}\frac{K'_\ell(w)}{K_\ell(w)}  \right]^{-1 } A_1  \label{Amplitudes_A1_B1_F1_G1}
.\end{align}

Now one has a full expression for the mode's fields. The next step is to find the number of photons $\braket{n}$ per $\mu m$. Once one calculates $g$ for classical field, the coupling constant is quantitatively retrieved \textit{from the classical-field calculation} by $\abs{alpha}=\frac{\abs{g}}{\abs{\beta}}=\frac{\abs{g}}{\sqrt{\braket{n}}}$. In practice, one can choose to normalize $A_1$ per photon so that $\braket{n}=1$ results (just for the simplicity of the calculation, not for that actual physical case) in the direct form $\alpha=g$.
 
For a propagating mode, the energy is time-stationary and azimutally uniform, so only the radial distribution requires calculation, at a given time. I choose the time of maximal $E_z$, along the axis $\phi=0$. Thus, one can ignore field components that nullify along the axis of $\phi=0$, or those with a temporal phase shift $i$, since their quarter-cycle shift nullifies when $E_z$ maximal. \\
The Field components other than $E_z$ are
\begin{align}
E_r&=-\frac{i\beta}{k_0^2n^2-\beta^2}\left[ \partial_r E_z +\frac{\mu_0\omega}{\beta r} \partial_\phi H_z\right] ,\,&\quad \text{out-of-phase in time or $\phi$}\\
E_\phi&=-\frac{i\beta}{k_0^2n^2-\beta^2}\left[ \frac{1}{r}\partial_\phi E_z  \boxed{-\frac{\mu_0\omega}{\beta } \partial_r H_z}\right] ,\,&\quad i\partial_r H_z \text{ is in phase}\\
H_r&=-\frac{i\beta}{k_0^2n^2-\beta^2}\left[ \boxed{\partial_r H_z } -\frac{\varepsilon_0 n^2 \omega}{\beta r} \partial_\phi E_z \right] ,\,&\quad i\partial_rH_z \text{ is in phase}\\
H_\phi&=-\frac{i\beta}{k_0^2n^2-\beta^2}\left[ \frac{1}{r}\partial_\phi H_z  +\frac{\varepsilon_0 n^2 \omega}{\beta } \partial_r E_z \right] ,\,&\quad i\partial_rH_z \text{out-of-phase in time or $\phi$}\\
H_z \, &\text{, see eq. \eqref{H_z_round_fiber}}&\quad  H_z \text{ is out-of-phase temporally}
.\end{align}
The in-phase components, spatially \textit{and} temporally are boxed. The others do not contribute. 
It is convenient to express the energy in terms of $\vec{E}$ and $\round{\mu_0\omega \vec{H}}$, since the factor $\mu_0\omega$ comes either from the above ratios or from eq. \eqref{Amplitudes_A1_B1_F1_G1}. The energy is
\begin{equation}
U=\half \int_{space}\round{\vec{E}\cdot \vec{D}+ \vec{B}\cdot \vec{H}}=\half \int_{space}\round{ \varepsilon_0 n^2 \abs{\vec{E}}^2 +  \mu_0\round{\frac{1}{\mu_0\omega}}^2\abs{\mu_0\omega\vec{H}}^2}.
\end{equation}
Using the relations $\varepsilon_0=\round{\mu_0 c^2}^{-1}$, one can write
\begin{equation}
U=\half \varepsilon_0 \int_{space}\round{  n^2 \abs{\vec{E}}^2 +  \round{\frac{c}{\omega}}^2\abs{\mu_0\omega\vec{H}}^2}
.\end{equation}
To normalize the fiber mode across some volume, we choose a fiber length $L\,[\mu m]$, with periodic boundary conditions, to allow for a unidirectional mode. This simplification can be easily taken into account in the cavity design, using the cavity effective length an any particular geometry. For example, the mode effective volume would be larger by a factor of $\sqrt{2}$ for a cavity encapsulated between two mirrors. I assume that the energy distribution is independent of $\phi$. \\
Using these fields, one can require that $A_1$ normalizes the energy to that of one photon,
\begin{align}
U&=A_1^2 \cdot L\cdot \varepsilon_0 \int_0^{2\pi}d\theta\int_{r=0}^\infty \round{  n^2 \round{E_z^2+E_\phi^2} +  \round{\frac{c}{\omega}}^2 \round{\mu_0\omega H_r}^2}rdr \overset{!}{=}\hbar\omega
,\end{align}
where one assumes the fields above were initially scaled according to eq. \eqref{Amplitudes_A1_eq_1}. Thus, $A_1$ is given by 
\begin{align}
A_1&=\sqrt{\frac{\hbar\omega}{2\pi\varepsilon_0 L}}\left[ \int_{r=0}^\infty \round{  n^2 \round{E_z^2+E_\phi^2} +  \round{\frac{c}{\omega}}^2 \round{\mu_0\omega H_r}^2}rdr \right]^{-\half}  \label{A_1_amplitude_per_1_photon}
.\end{align}
This integral is evaluated for the inner and outer segments, $\int_0^a$ and $\int_a^\infty$ using $\round{J_1(),n_{core}}$ and $\round{K_1(),n_{clad}=1}$ , respectively. \\
At this point one has the classical field of an $HE_{11}$ fiber mode with an average energy of one photon. To evaluate the coupling constant, one only needs to calculate $g$ via its definition in refs. \cite{feist_quantum_2015,echternkamp_ramsey_type_2016} 
\begin{equation}
g=\frac{q}{2\hbar\omega}\int_0^L{E_z\round{r,\phi=0,z,t(z)}}
,\end{equation}
for the electron trajectory $\round{z,t(z)}$ (see main text). Only $E_z(r>a)$ is relevant to accelerate/decelerate an electron in vacuum. 
\begin{align}
E_z\round{r>a,\phi,z,t}  &= B_1 K_1\round{w\frac{r}{a}}e^{i\round{\omega t-\beta z}}\\
B_1  &=  A_1\frac{J_1(u)}{K_1(w)}\\
.\end{align}
The maximal relevant field is, available for electron coupling right at the fiber edge is
\begin{align}
E_z\round{a^+,0,0,0}=&B_1 K_1(w)\\
.\end{align}
For a phase-matched interaction the electron experience a time-independent field along its path $\round{z,t(z)}$, so 
\[ E\round{r,\phi,z,t(z)}=E\round{r,\phi,0,0}. \]
Thus, the maximal PINEM interaction per photon, is on the surface of the fiber, under conditions of full-phase-matching is 
\begin{equation}
g_{\text{per photon}}^{\text{max PINEM}} = \frac{1}{2\hbar\omega}\int_0^L{E_z\round{a^+,0,z,t(z)}dz}  =  \frac{q E_z \round{a^+,0,0,0} }{2\hbar\omega}L
.\end{equation}
This is, quantitatively, the maximum electron-photon coupling
\begin{equation}
\boxed{\alpha_{\text{max}}=\frac{g}{\sqrt{\braket{n}}}=\frac{q E_z \round{a^+,0,0,0} }{2\hbar\omega}L} 
.\end{equation}
One should note that the maximal coupling scales with the cavity length as 
\[ 
\alpha_{\text{max}}\propto \sqrt{L}
\], 
since the volume normalization included in $A_1$, scales as $1/\sqrt{L}$.

\begin{figure}[H]
\centering
\includegraphics[width=0.49\textwidth]{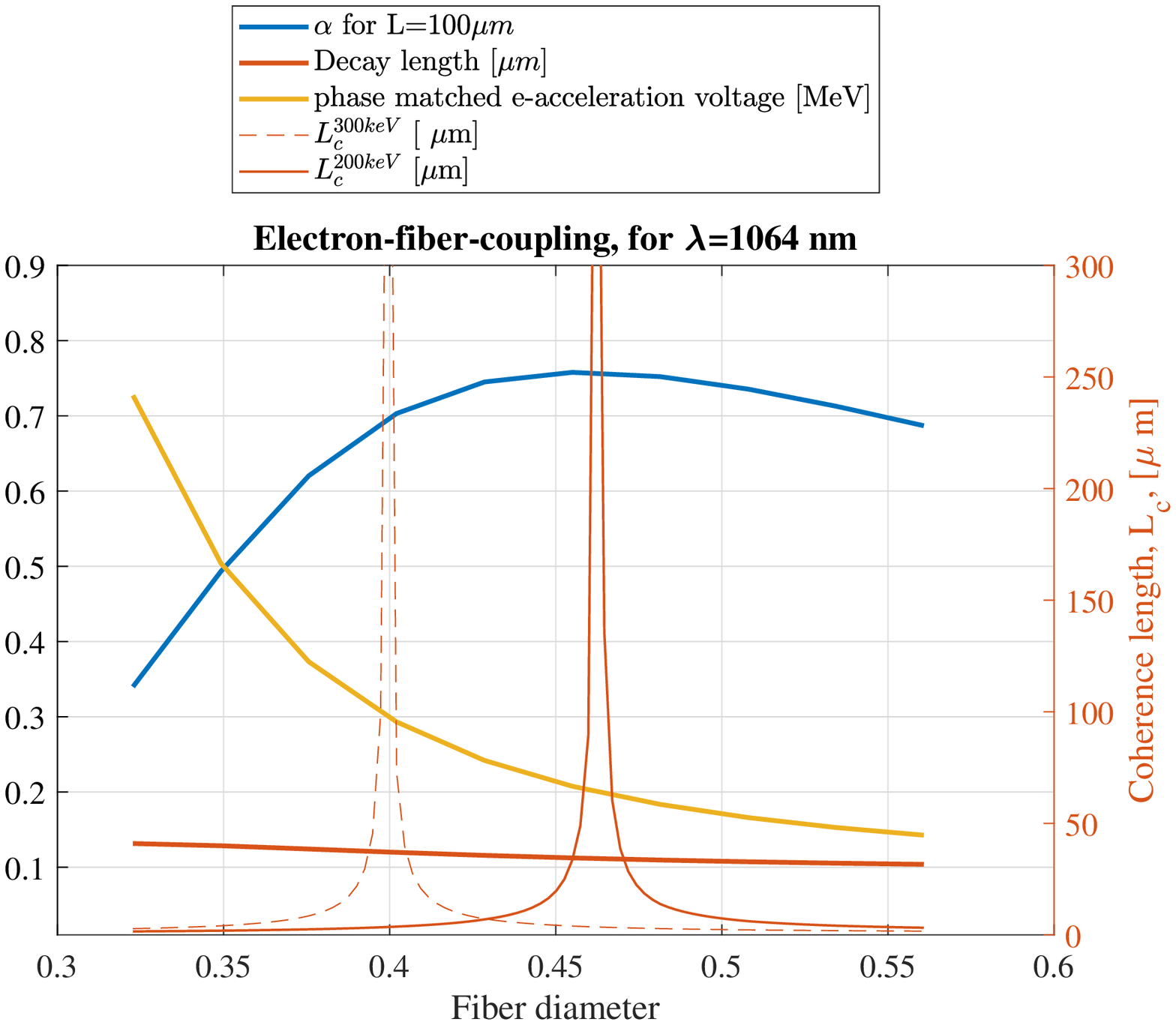}
\includegraphics[width=0.49\textwidth]{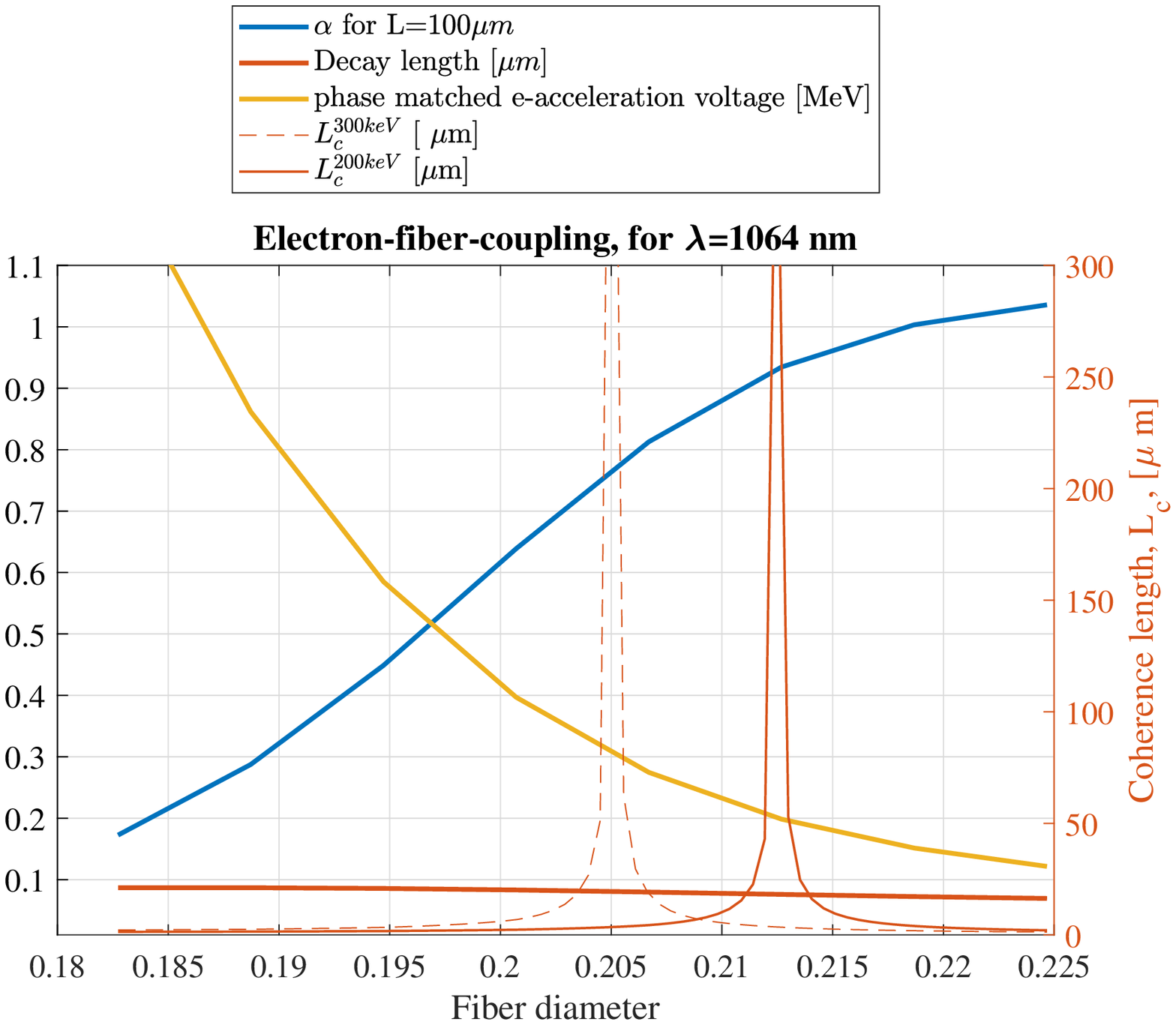}
\caption{Relevant properties for the coupling of relativistic electrons to Si$_3$N$_4$-core (left) and Si-core (right) in vacuum. The waveguide width is chosen based on the optimal phase matching condition - note the divergence of $L_c$ for 200 keV electrons  at a diameter of 463 nm and 213 nm in Si$_3$N$_4$ and Si, respectively. The energy selectivity for these parameters is shown in figure 4b in the main text. This regime determines the coupling constant, and the characteristic decay length of the field out of the fiber ($e^{-1} $ distance). Different diameters are optimized for either slower or faster electrons.}
\label{Full_Coupling_properties_Si3N4_1064nm}

\end{figure}

\begin{figure}[H]
\centering
\includegraphics[width=10cm]{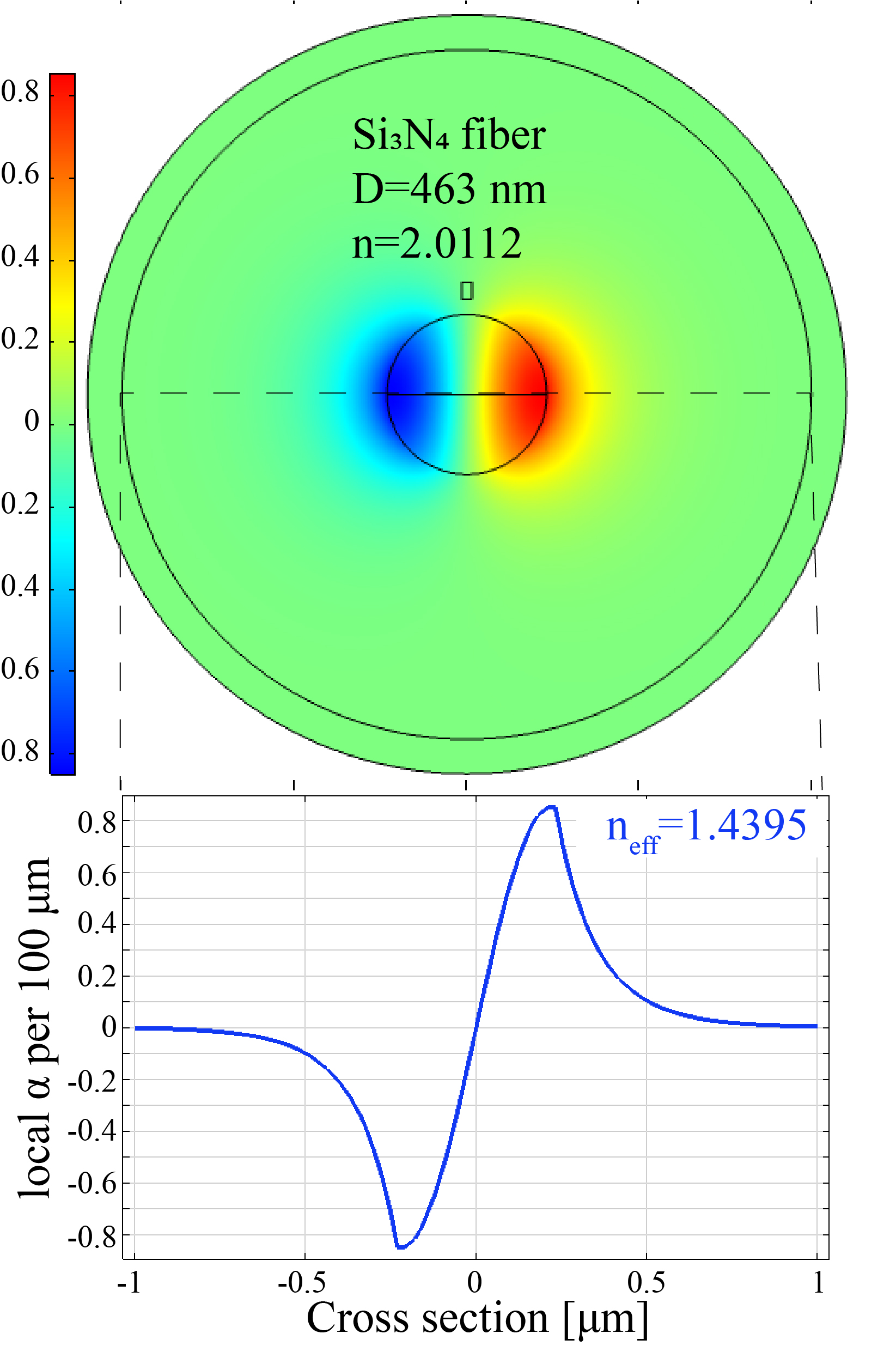}
\caption{(top) Distribution of the field component $E_z$ in a step index fiber of a Si$_3$N$_4$ in vacuum. The inner circle is the core, and the outer two circles form the boundaries for the numerical box. (bottom) a cross-section of the above colormap. The field component $E_z$ is normalized to represent the field per photon, in a fiber of length $L=100 \mu m$ with periodic boundary conditions, multiplied by $\frac{q}{2\hbar\omega_0}L$. That is, it is the coupling constant $\alpha$ for a electron that travels in a fixed distance parallel to such a fiber. The calculation is done for vacuum wavelength $\lambda_0=1064 nm$. The simulations verify the analytic calculation of the mode properties, used to evaluate the coupling constant. }
\label{COMSOL_463_nm_Si3N4}

\end{figure}

For a step index fiber of Si$_3$N$_4$-core in vacuum having length of $100\, \mu m$ with periodic boundary conditions, I calculated the coupling properties vs. the fiber diameter (figure \ref{Full_Coupling_properties_Si3N4_1064nm}). The calculated properties are the optimal coupling, the distance for $e^{-1}$ decay of the field, the acceleration voltage for phase-matched electrons, and on the right axis, the coherence lengths for acceleration voltages of 200 keV and 300 keV. The calculations for the fields of the electromagnetic mode, and the normalization terms were varified using COMSOL Multiphysics${}^{\text{\textregistered}}$, shown on figure \ref{COMSOL_463_nm_Si3N4}.

\newpage
\begin{appendices}
\section{Assisting derivations}
    \begin{itemize}
   
\begin{comment}
\item
\textbf{Derivation of eq. \eqref{electron_light_coherent_State_overlap}}
\begin{align}
    \left<{E_{j'-\alpha'},\alpha'} | {E_{j-\alpha},\alpha}\right>
    &=\sum_{n,n'=0}^{\infty}{e^{-{\abs{\alpha}^2}/{2}}e^{-{\abs{\alpha'^*}^2}/{2}}\frac{\alpha^n{\alpha'}^{n'}}{\sqrt{n!n'!}} \underbrace{\left<{E_{j'-n'},n'} | {E_{j-n},n}\right>}_{=\delta_{n,n'}\delta_{j'-n',j-n}=\delta_{n,n'}\delta_{j',j}}}\\
    &=\sum_{n=0}^{\infty}e^{-{\abs{\alpha}^2}/{2}}e^{-{\abs{\alpha'}^2}/{2}}\frac{\round{\alpha \alpha'^*}^{n}}{n!}\delta_{j,j'}\\
    &=\delta_{j,j'}e^{-{\abs{\alpha}^2}/{2}}e^{-{\abs{\alpha'}^2}/{2}} \sum_{n=0}^{\infty}\frac{\round{\alpha \alpha'^*}^{n}}{n!}
.\end{align}
So,
\begin{align}
    \left<{E_{j'-\alpha'},\alpha'} | {E_{j-\alpha},\alpha}\right>
    &=\delta_{j,j'}e^{-{\abs{\alpha}^2}/{2}}e^{-{\abs{\alpha'}^2}/{2}} e^{{{\alpha{\alpha'}^*}}}\\
    &=\delta_{j,j'}e^{-{\abs{\alpha-\alpha'}^2}/{2}}\label{Overlap_E_alpha_E_alpha_tag}
.\end{align}
In the specific case, where $\alpha'=\alpha$
\begin{align}
    \left<{E_{j'-\alpha},\alpha} | {E_{j-\alpha},\alpha}\right>&=\delta_{j,j'}\label{Overlap_E_alpha_E_alpha}
\end{align}
\end{comment}

\item \textbf{Explicit derivation of BCH}\\
According to BCH, for operators that obey $[X,Y]=const$
\begin{align}
    &e^{X+Y}=e^{X}e^{Y}e^{-\half [X,Y ]}\\
    &e^{X+Y}=e^{Y}e^{X}e^{+\half [X,Y ]}
,\end{align}
where I just stressed the importance of the sign.
For the displacement operator,
\begin{align}
X=\alpha \hat{b} a^\dagger, \, Y=-\alpha^* \hat{b}^\dagger a, \, \half [X,Y]=\half \left[ \alpha \hat{b} a^\dagger,-\alpha^* \hat{b}^\dagger a\right] =\half\abs{\alpha}^2\\
.\end{align}
which means that 
\begin{align}
    D_{\round{\hat{b}\alpha}}&=e^{\alpha \hat{b} a^\dagger -\alpha^* \hat{b}^\dagger a }\\
    &=e^{\mathbf{-}\half \abs{\alpha}}e^{\alpha \hat{b} a^\dagger} e^{-\alpha^* \hat{b}^\dagger a }\\
    &=e^{\mathbf{+}\half \abs{\alpha}} e^{-\alpha^* \hat{b}^\dagger a }e^{\alpha \hat{b} a^\dagger}\label{BCH_explicitly_for_Displacement_Operator}
.\end{align}

\item 
\textbf{Approximating $\round{N+\ell}!\approx N!\round{N}^{\ell}$}
using Stirling's Formula, $z!\approx \sqrt{2\pi z}\round{\frac{z}{e}}^z$. Taking the natural logarithm result in  \[
ln\round{z!}\approx z ln\round{z}-z+\half ln\round{2\pi z}
.\] 
I will need to convert $ln\round{N+\ell}$ to a form with $ln\round{N}$, so explicitly
 \begin{align}
 ln\round{N+\ell}&=ln\round{N\round{1+\frac{\ell}{N}}}\\
 &=ln\round{N}+ln\round{1+\frac{\ell}{N}}
, \end{align}
which by Taylor expantion to the first order ,provides 
\begin{equation}
 ln\round{N+\ell}
 = ln\round{N}+\frac{\ell}{N}+\mathcal{O}\round{\frac{\ell}{N}}^2 \label{ln_N_plus_ell_Taylor}
.\end{equation}

Expanding the factorial $\round{N+\ell}!$ according to the above,
\begin{align}
 ln\round{\round{N+\ell}!}&\approx \round{N+\ell} ln\round{N+\ell}-\round{N+\ell}+\half ln\round{2\pi \round{N+\ell}}\\
 &\overset{eq.\, \eqref{ln_N_plus_ell_Taylor}}{\approx} \underbrace{ N ln\round{N}-N+\half ln\round{2\pi N}}_{ln\round{N!}} +\underbrace{\round{N+\ell}\frac{\ell}{N}}_{\ell+\frac{\ell^2}{N}}+\ell\cdot  \underbrace{ln\round{N+\ell}}_{ln \round{N}+\frac{\ell}{N}}-\ell
.\end{align}
Neglecting terms of order $\round{\frac{\ell}{N}}^2$, this resuls in 
\begin{equation*}
ln\round{\round{N+\ell}!}=ln\round{N!}+\ell ln\round{N} +2\frac{\ell^2}{N}
.\end{equation*}
Within a correction of $e^{\round{\frac{\ell}{N}}^2}$, one gets, \[
\round{N+\ell}! = N!\round{N}^{\ell} e^{2\frac{\ell^2}{N}} +\mathcal{O}\round{\frac{\ell}{N}}^2.\]
For simplification, in the approximation of $2 \ell^2 \ll N$, this leads to the final result
\begin{align}
\round{N+\ell}!&\approx N!\round{N}^{\ell} \label{Approximating_N_pule_l_factorial}
.\end{align}
Which we use to simplify the factorial terms in the electron-photon coupling and electron-electron coupling.

\end{itemize}

\end{appendices}

\bibliography{Supplementary_Material_v7}

\begin{thebibliography}{1}

\bibitem{feist_quantum_2015}
Armin Feist, Katharina~E. Echternkamp, Jakob Schauss, Sergey~V. Yalunin, Sascha
  Sch\"afer, and Claus Ropers.
\newblock Quantum coherent optical phase modulation in an ultrafast
  transmission electron microscope.
\newblock {\em Nature}, 521(7551):200--203, May 2015.

\bibitem{scully_quantum_1997}
Marlan~O. Scully and M.~Suhail Zubairy.
\newblock {\em Quantum {Optics}}.
\newblock Cambridge University Press, September 1997.

\bibitem{Mandel_Wolf_optical_1995}
Leonard Mandel and Emil Wolf.
\newblock {\em Optical {Coherence} and {Quantum} {Optics}}.
\newblock Cambridge University Press, September 1995.

\bibitem{garcia_de_abajo_multiple_2013}
F.~Javier Garc{\'i}a~de Abajo.
\newblock Multiple excitation of confined graphene plasmons by single free
  electrons.
\newblock {\em ACS Nano}, 7(12):11409--11419, December 2013.

\bibitem{echternkamp_ramsey_type_2016}
Katharina~E. Echternkamp, Armin Feist, Sascha Sch{\"a}fer, and Claus Ropers.
\newblock Ramsey-type phase control of free-electron beams.
\newblock {\em Nature Physics}, 12(11):1000--1004, November 2016.

\bibitem{yeh_guided_wave_1987}
C.~Yeh.
\newblock Guided-wave modes in cylindrical optical fibers.
\newblock {\em IEEE Transactions on Education}, E-30(1):43--51, February 1987.

\bibitem{chen_foundations_2006}
Chin-Lin Chen.
\newblock {\em Foundations for {Guided}-{Wave} {Optics}}.
\newblock John Wiley \& Sons, September 2006.
\newblock Google-Books-ID: LxzWPskhns0C.

\end{thebibliography}

\end{document}